\documentclass[10pt]{iopart}
\usepackage{graphicx}
\usepackage{iopams}

\newcommand{\sgn}{\mathop{\mathrm{sgn}}\nolimits}

\newcommand{\abs}[1]{|#1|}
\newcommand{\binom}[2]{\left(\begin{array}{c}#1\\#2\end{array}\right)}
\newcommand{\vect}[1]{\bi{#1}}
\newcommand{\journaltitle}[1]{\textit{#1}}
\newcommand{\booktitle}[1]{\textit{#1}}
\newcommand{\volume}[1]{\textbf{#1}}
\newcommand{\dlE}{\epsilon}
\newcommand{\dlk}{\kappa}
\newcommand{\dlx}{\bar\xi}
\newcommand{\dldx}{\xi}
\newcommand{\dly}{\eta}
\newcommand{\dlz}{\zeta}
\newcommand{\dlmu}{\nu}

\newcommand{\Imag}{\mathop{\mathrm{Im}}\nolimits}
\begin{document}
\jl{1}

\title{Correlations in a confined magnetized free-electron gas}
\author{M M Kettenis and L G Suttorp}
\address{Instituut voor Theoretische Fysica, Universiteit van Amsterdam, 
Valckenierstraat 65, 1018 XE Amsterdam, The Netherlands}

\begin{abstract}
Equilibrium quantum statistical methods are used to study the pair correlation 
function for a magnetized free-electron gas in the presence of a hard wall 
that is parallel to the field. With the help of a path-integral technique and 
a Green function representation the modifications in the correlation function 
caused by the wall are determined both for a non-degenerate and for a 
completely degenerate gas. In the latter case the asymptotic behaviour of the 
correlation function for large position differences in the direction parallel 
to the wall and perpendicular to the field, is found to change from Gaussian 
in the bulk to algebraic near the wall.
\end{abstract}   

\pacs{05.30.Fk,75.20.-9} 

\submitted 

\maketitle

\section{Introduction}
As has been known since Bohr, van Leeuwen and Landau \cite{BOH:1911}, edge 
effects play an important role in the physics of magnetized 
charged-particle systems in equilibrium. In particular, the diamagnetic 
response of these systems in the quantum regime is determined by electric 
currents flowing near the walls. The profiles of these currents, and of the 
closely related particle density, for a non-interacting magnetized electron 
gas near a hard wall parallel to the magnetic field have been investigated 
in detail \cite{OHT:197397}, \cite{JOH:1994} -- \cite{KES:1999}. Much less is 
known about the profiles in an interacting magnetized electron gas. 

Equally important for a physical understanding of the properties of an 
equilibrium quantum system are the correlation functions. For positions in the 
bulk of the system these have been studied extensively, both for a 
non-interacting magnetized electron gas and for its interacting counterpart. 
For the non-interacting gas the bulk pair correlation function can be 
determined analytically both for dilute systems at high temperatures and for 
dense low-temperature systems, in which quantum degeneracy effects are 
important \cite{ISI:1971}. For the interacting electron gas information on the 
behaviour of the bulk correlation functions is more difficult to obtain. Even 
for the non-magnetized case these functions have surprising properties. In 
fact, it has been demonstrated that the bulk correlation functions of the 
interacting electron gas possess slowly decaying tails with an algebraic 
dependence on the position difference \cite{ALA:198897}. For the magnetized 
interacting gas similar methods have been employed to prove the existence of 
analogous algebraic tails, albeit with a different exponent \cite{COR:199798}.

The correlation functions are expected to change near a hard wall. For a 
non-magnetized free-particle system these changes are easily determined by 
using a reflection principle \cite{ROEP:1996}. The problem becomes a lot more 
complicated when either interactions between the particles or a magnetic field 
or both are incorporated. In a recent paper \cite{AQU:1999} the interactions 
between the charged particles have been taken into account in a model system 
consisting of two quantum charges immersed in a classical plasma confined by a 
wall. An algebraic tail in the pair correlation function of the quantum 
particles near the wall was found. However, the exponent governing the 
algebraic tail turned out to be different from that of the bulk correlation 
functions discussed above. The result corroborates earlier findings based on 
linear response arguments \cite{JAN:1985}. The influence of a magnetic field 
on the surface correlation functions in the adopted model system remains to be 
studied. 

As the influence of a wall on the correlations in magnetized quantum systems 
is not yet fully understood, it appears to be useful to try and investigate 
the correlation functions for the relatively simple case of a non-interacting 
magnetized electron gas in the presence of a hard wall. In the following we 
present some new results for this system. In particular, we will analyze the 
correlations for the strongly degenerate case of high density and low 
temperature, where the influence of Fermi statistics is important. Both strong 
and weak fields will be considered, so that the number of filled Landau levels 
can vary considerably.

The paper is organized as follows. We start with two sections that serve to 
prepare the ground. In section 2 we define the relevant correlation functions 
for a system of independent particles and discuss their relation to the 
one-particle Green functions. The pair correlation function in the bulk is 
considered in section 3, where the influence of the magnetic field on the 
correlations is determined both analytically and numerically. After these 
preparatory sections we start considering the influence of the wall in section 
4. In that section we use the so-called `path-decomposition expansion', which 
follows from a path-integral formulation, to determine the lowest-order 
corrections in the correlation functions at positions in the transition 
region, where the presence of the wall starts to be felt. An alternative way 
to determine these corrections is based on an eigenfunction expansion of the 
Green function, which is the subject of section 5. The asymptotic form of the 
correlation functions for large position differences is established in section 
6, separately for directions parallel with and transverse to the magnetic 
field. In section 7 the correlation functions for positions close to the wall 
are studied, again for both directions. In the final section 8 some 
conclusions will be drawn. 

\section{Correlations}

The equilibrium quantum statistical properties of a system of independent 
particles are determined by the temperature Green function
\begin{equation}
\label{2.1}
G_\beta(\vect{r},\vect{r}')=\langle \vect{r}|\rme^{-\beta 
H}|\vect{r}'\rangle=\sum_n\, \rme^{-\beta E_n}\, \psi_n(\vect{r})\,
\psi_n^*(\vect{r}').
\end{equation}
Here $\beta$ is the inverse temperature, and $\psi_n(\vect{r})$ and $E_n$ 
the eigenfunctions and eigenvalues of the one-particle Hamiltonian $H$, which 
is assumed to be independent of the spin of the particles. The reduced 
single-particle density matrix $\rho_{\beta,\mu}(\vect{r}, \vect{r}')$ of 
such a system at inverse temperature $\beta$ and chemical potential $\mu$ is 
found by incorporating the effects of quantum degeneracy. For Fermi-Dirac 
particles one has
\begin{equation}
\label{2.2}
\rho_{\beta,\mu}(\vect{r}, \vect{r}')=2\sum_n 
\frac{1}{1+\rme^{\beta(E_n-\mu)}} \,
\psi_n(\vect{r})\,\psi_n^*(\vect{r}')
\end{equation}
where the spin degeneracy has been taking into account. The local particle 
density $\rho_{\beta,\mu}(\vect{r})$ is the diagonal part of (\ref{2.2}).

For a completely degenerate system at zero temperature the reduced 
single-particle density matrix becomes
\begin{equation}
\label{2.3}
\rho_{\mu}(\vect{r}, \vect{r}')=2\sum_n \,\theta(\mu-E_n)\,
\psi_n(\vect{r})\, \psi_n^*(\vect{r}')\equiv G_{\mu}(\vect{r}, \vect{r}')
\end{equation}
with $\theta$ the step function. The diagonal part gives the local particle 
density $\rho_{\mu}(\vect{r})$ of the completely degenerate system. The 
$\mu$-dependent Green function, as defined here, is obtained from the 
temperature Green function by an inverse Laplace transform \cite{SOW:1951}
\begin{equation}
  \label{2.4}
  G_\mu(\vect{r},\vect{r}') =
  \frac{1}{2\pi\rmi}\int_{c-\rmi\infty}^{c+\rmi\infty} \rmd\beta\,
  \rme^{\beta\mu}\, \frac{2}{\beta}\, G_\beta(\vect{r},\vect{r'}).
\end{equation}
with $c>0$.                                     

The $n$-particle reduced density matrix $\rho^{(n)}_{\beta,\mu}(\vect{r}, 
\vect{r}')$ follows from its one-particle counterpart by a symmetrized 
factorization:
\begin{equation}
  \label{2.5}
  \rho^{(n)}_{\beta,\mu}(\vect{r}^n, \vect{r}'^n) = \sum_{\pi \in \mathcal{S}_m}
  \epsilon^\pi \prod_{j=1}^n \rho_{\beta,\mu}(\vect{r}_j, \vect{r}'_{\pi(j)}).
\end{equation}
Here the sum is taken over all permutations of the $n$ position vectors, with 
$\epsilon^\pi$ the sign of the permutation. The structure of the $n$-particle 
reduced density matrix has been analyzed quite generally for a system of 
interacting particles by using a path-integral formalism \cite{GIN:1971}. The 
factorization property for a system of independent particles then follows as a 
special case. The argument is not changed by incorporating an external 
magnetic field and a hard wall that confines the system.

For the completely degenerate case a relation similar to (\ref{2.5}) holds for 
$\rho^{(n)}_{\mu}(\vect{r}^n, \vect{r}'^n)$ . In particular, the diagonal part 
of the two-point 
reduced density matrix at zero temperature is
\begin{equation}
\label{2.6}
\rho^{(2)}_{\mu}(\vect{r} \vect{r}',\vect{r} \vect{r}') =
  G_\mu(\vect{r}, \vect{r}) G_\mu(\vect{r}', \vect{r}') -
  G_\mu(\vect{r}, \vect{r}') G_\mu(\vect{r}', \vect{r}).
\end{equation}
Often it is convenient to introduce the two-point correlation function 
\begin{equation}
\label{2.7}
  g(\vect{r},\vect{r}') = 
\frac{\rho^{(2)}_{\mu}(\vect{r}\vect{r}',\vect{r}\vect{r}')}
{\rho_{\mu}(\vect{r})\rho_{\mu}(\vect{r}')} - 1
=-\frac{|G_{\mu}(\vect{r},\vect{r}')|^2}
{G_\mu(\vect{r}, \vect{r}) G_\mu(\vect{r}', \vect{r}')}
\end{equation}
In the following we will study this correlation function, and the influence of 
a magnetic field and a hard wall on its properties.

\section{Correlations in the bulk}

We consider a system of charged particles which move in a magnetic field 
directed along the $z$-axis. The interaction between the particles is 
neglected. To describe the magnetic field we adopt the Landau gauge, with 
vector potential $\vect{A}=(0,Bx,0)$.  The particles are confined to the 
half-space $x>0$ by a plane hard wall at $x=0$. 

For positions far from the wall the temperature Green function 
$G_{\beta}(\vect{r},\vect{r}')$ reduces to the bulk Green function 
$G_{\beta}^b(\vect{r},\vect{r}')$. The latter is given by \cite{SOW:1951}
\begin{eqnarray}
\label{3.1}
  \fl G^b_\beta(\vect{r},\vect{r}') =
  \frac{1}{\sqrt{2\pi\beta}}\,
  \frac{B}{4\pi\sinh(\beta B/2)}\,
  \exp\left[
    -\frac{B}{4\tanh(\beta B/2)}(\vect{r}_\perp-\vect{r}'_\perp)^2
  \right]
  \nonumber \\ \times
  \exp\left[
    \frac{\rmi B}{2}(x+x')(y-y')
  \right]\,
  \exp\left[-\frac{(z-z')^2}{2\beta}\right].
\end{eqnarray}
Units have been chosen such that the charge and the mass of the particles drop 
out, while $\hbar$ and $c$ have been put to 1 as well. From now on we will 
often measure distances in terms of the cyclotron radius $1/\sqrt{B}$. To that 
end we introduce the dimensionless variables $\dlx=\sqrt{B}(x+x')/2$, 
$\dldx=\sqrt{B}(x-x')$, $\dly=\sqrt{B}(y-y')$, and $\dlz=\sqrt{B}(z-z')$. 

The $\mu$-dependent Green function follows by inserting (\ref{3.1}) in 
(\ref{2.4}). One finds
\begin{eqnarray}
\label{3.2}
  \fl G^b_\mu(\vect{r},\vect{r}') =
  \frac{1}{2\pi\rmi}
  \int_{c-\rmi\infty}^{c+\rmi\infty} \rmd\beta\,
  \rme^{\beta B\nu}\,
  \frac{B}{2^{5/2}\,\pi^{3/2}\,\beta^{3/2}}
  \frac{1-{q_0}^2}{q_0}
  \nonumber \\ \times
  \exp\left[\rmi\,\dlx\dly - \frac{1+{q_0}^2}{8q_0} (\dldx^2 + \dly^2)\right]
  \exp\left[-\frac{\dlz^2}{2\beta B}\right]
\end{eqnarray}
with $q_0 = \tanh(\beta B/4)$. The chemical potential is measured in terms of 
the energy difference between adjacent Landau levels, by employing the 
dimensionless variable $\nu=\mu/B$. Let us use now the generating-function 
identity \cite{MOS:1966}
\begin{equation}
\label{3.3}
  \exp\left[-\frac{(1-q_0)^2}{8q_0}\dldx^2\right] =
  \frac{4q_0}{(1+q_0)^2} \sum_{n=0}^{\infty}
  L_n\left(\frac{\dldx^2}{2}\right)\left(\frac{1-q_0}{1+q_0}\right)^{2n}
\end{equation}
to express the exponential function in terms of Laguerre polynomials. After 
substitution of $(1-q_0)/(1+q_0)=\rme^{-\beta B/2}$ this gives
\begin{eqnarray}
  \label{3.4}
  \fl G^b_\mu(\vect{r},\vect{r}) =
  \frac{B^{3/2}}{\sqrt{2}\,\pi^{3/2}}
  \exp\left(\rmi\,\dlx\dly - \frac{\dldx^2+\dly^2}{4}\right)
  \sum_{n=0}^\infty L_n \left(\frac{\dldx^2+\dly^2}{2}\right)
  \nonumber \\ \times
  \frac{1}{2\pi\rmi}
  \int_{c'-\rmi\infty}^{c'+\rmi\infty} \rmd s\,
  \rme^{s[\nu-(n+1/2)]}\,
  s^{-3/2}\,
  \exp\left(-\frac{\dlz^2}{2s}\right)
\end{eqnarray}
where we have set $s = \beta B$ and $c'=cB$. The sum over $n$ can be 
interpreted as a sum over Landau levels.

The inverse Laplace transform in (\ref{3.4}) can be found in 
\cite{ERD:1954}:
\begin{equation}
  \label{3.5}
  \frac{1}{2\pi\rmi} \int_{c-\rmi\infty}^{c+\rmi\infty} \rmd s\, 
\rme^{st}\,
  s^{-3/2}\,\rme^{-a/s} =
  \frac{1}{\sqrt{\pi a}}\, \sin\big(2 \sqrt{at}\, \big)\, \theta(t)
\end{equation}
for $a>0$ and $c>0$. Use of this identity in (\ref{3.4}) results in the 
following expression for the $\mu$-dependent Green function in the bulk:
\begin{eqnarray}
  \label{3.6}
  \fl G^b_\mu(\vect{r},\vect{r}') =
  \frac{B^{3/2}}{\pi^2}
  \exp\left(\rmi\,\dlx\dly - \frac{\dldx^2+\dly^2}{4}\right)\nonumber\\
\times  {\sum_n}' L_n \left(\frac{\dldx^2+\dly^2}{2}\right)
  \frac{\sin\big(\sqrt{2[\nu-(n+1/2)]}\, \dlz\big)}{\dlz}.
\end{eqnarray}
The prime indicates that the summation is only over those values of $n$ for 
which $\dlmu-(n+1/2)$ is positive, i.e. over the Landau levels that are at 
least partially filled. The bulk density follows as the diagonal part of 
(\ref{3.6})
\begin{equation}
\rho^b_{\mu}= G^b_\mu(\vect{r},\vect{r}) =
  \frac{\sqrt{2}\, B^{3/2}}{\pi^2}
{\sum_n}' \sqrt{\nu-(n+1/2)}.
\label{3.7}
\end{equation}
The two-particle correlation function is found upon substitution of 
(\ref{3.6}) and (\ref{3.7}) in (\ref{2.7}). 

The expressions (\ref{3.6}) and (\ref{3.7}) are particularly useful for strong 
fields when only a few Landau levels are occupied. An alternative expression, 
which is useful for weak fields only, has been derived in \cite{ISI:1971}. To 
study the limit $B\rightarrow 0$ in (\ref{3.6}), we return to the original 
variables, since measuring the distances in terms of the cyclotron radius, or 
the chemical potential in terms of the energy difference between adjacent 
Landau levels, becomes meaningless for vanishing magnetic fields. The number 
of terms in the sum becomes large, as the upper limit is inversely 
proportional to $B$ at fixed $\mu$. Furthermore, the argument of the Laguerre 
polynomial gets small for fixed $x-x'$ and $y-y'$. Hence, we can use the 
asymptotic form of the Laguerre polynomials \cite{MOS:1966}
\begin{equation}
\label{3.8}
  L_n(u) \approx \rme^{u/2} J_0\big(\sqrt{2(2n+1)u}\, \big)
\end{equation}
which is valid for $u/(n+1/2)^{1/3}\ll 1$. Use of this approximation gives
\begin{eqnarray}
\label{3.9}
  \fl G^b_\mu(\vect{r},\vect{r}') \approx
  \frac{B^{3/2}}{\pi^2}\,
  \exp\left[\frac{\rmi B (x+x')(y-y')}{2}\right]
  {\sum_n}' \;J_0\big(\sqrt{2B(n+1/2)} \, 
\abs{\vect{r}_\perp-\vect{r}'_\perp}\big)
  \nonumber\\ \times
  \frac{\sin\big[\sqrt{2[\mu - B(n+1/2)]}\, (z-z')\big]}{\sqrt{B}\, (z-z')}
\end{eqnarray}
for small magnetic fields.  The subscripts $\mbox{}_{\perp}$ denote the 
transverse parts of the position vectors, which follow by projection on the 
$xy$-plane. 

If $B$ approaches zero, the number of Landau levels becomes very large, and 
their spacing becomes very small.  Therefore it is permitted to replace the 
summation over Landau levels in (\ref{3.9}) with an integral. In the limit of 
vanishing $B$ we get
\begin{equation}
\label{3.10}
  \fl G^b_\mu(\vect{r},\vect{r}') =
  \frac{1}{\pi^2}
  \int_0^\mu \rmd t\, J_0\big(\sqrt{2 t}\, 
\abs{\vect{r}_\perp-\vect{r}'_\perp}\big)\,
  \frac{\sin \big[\sqrt{2(\mu - t)}\, (z-z')\big]}{z-z'}.
\end{equation}
With the help of the identity \cite{ERD:1954a}
\begin{eqnarray}
\label{3.11}
  \fl \int_0^a \rmd x\, x^{\nu+1}\, \sin\big(b\sqrt{a^2-x^2}\, \big)\, 
J_\nu(x) \nonumber\\
=  \sqrt{\frac{\pi}{2}}\, a^{\nu+3/2}\, b\,(1+b^2)^{-\nu/2-3/4}\,
  J_{\nu+3/2}\big(a\sqrt{1+b^2}\, \big) 
\end{eqnarray}
for $\nu =0$, we arrive at
\begin{equation}
\label{3.12}
  G^b_\mu(\vect{r},\vect{r}') =
  \frac{2^{1/4}\,\mu^{3/4}}{\pi^{3/2}}\,
  \frac{1}{\abs{\vect{r}-\vect{r}'}^{3/2}}\,
  J_{3/2}\big(\sqrt{2\mu}\, \abs{\vect{r}-\vect{r}'}\big).
\end{equation}
Note that the right-hand side is an isotropic function of the position 
difference, as should be the case for a vanishing magnetic field. We can 
simplify it further by using the explicit form for the Bessel function
\begin{equation}
\label{3.13}
  J_{3/2}(u) = \sqrt{\frac{2}{\pi}} \, \frac{\sin u - u \cos u}{u^{3/2}}.
\end{equation}
The final result is
\begin{eqnarray}
  \label{3.14}
  \fl G^b_\mu(\vect{r},\vect{r}') =
  - \frac{1}{\pi^2 \abs{\vect{r}-\vect{r}'}^2}\nonumber\\
\times  \left[
    \sqrt{2\mu}\, \cos\big(\sqrt{2\mu}\, \abs{\vect{r}-\vect{r}'}\big) -
    \frac{1}{\abs{\vect{r}-\vect{r}'}}\, 
    \sin\big(\sqrt{2\mu}\, \abs{\vect{r}-\vect{r}'}\big)
  \right]
\end{eqnarray}
which is identical to what one gets by starting from the temperature Green 
function for the non-magnetized system
\begin{equation}
\label{3.15}
  G^b_\beta(\vect{r},\vect{r}') =
  \frac{1}{(2\pi\beta)^{3/2}}\,
  \exp\left[-\frac{(\vect{r} - \vect{r}')^2}{2\beta}\right]
\end{equation}
and applying (\ref{2.4}). The bulk density in the field-free case is 
$\rho_{\mu}^b=(2\mu)^{3/2}/(3\pi^2)$. The two-particle correlation function 
in the bulk follows upon inserting (\ref{3.14}) in (\ref{2.7})

\begin{figure}
  \begin{center}
    \includegraphics[height=6.5cm]{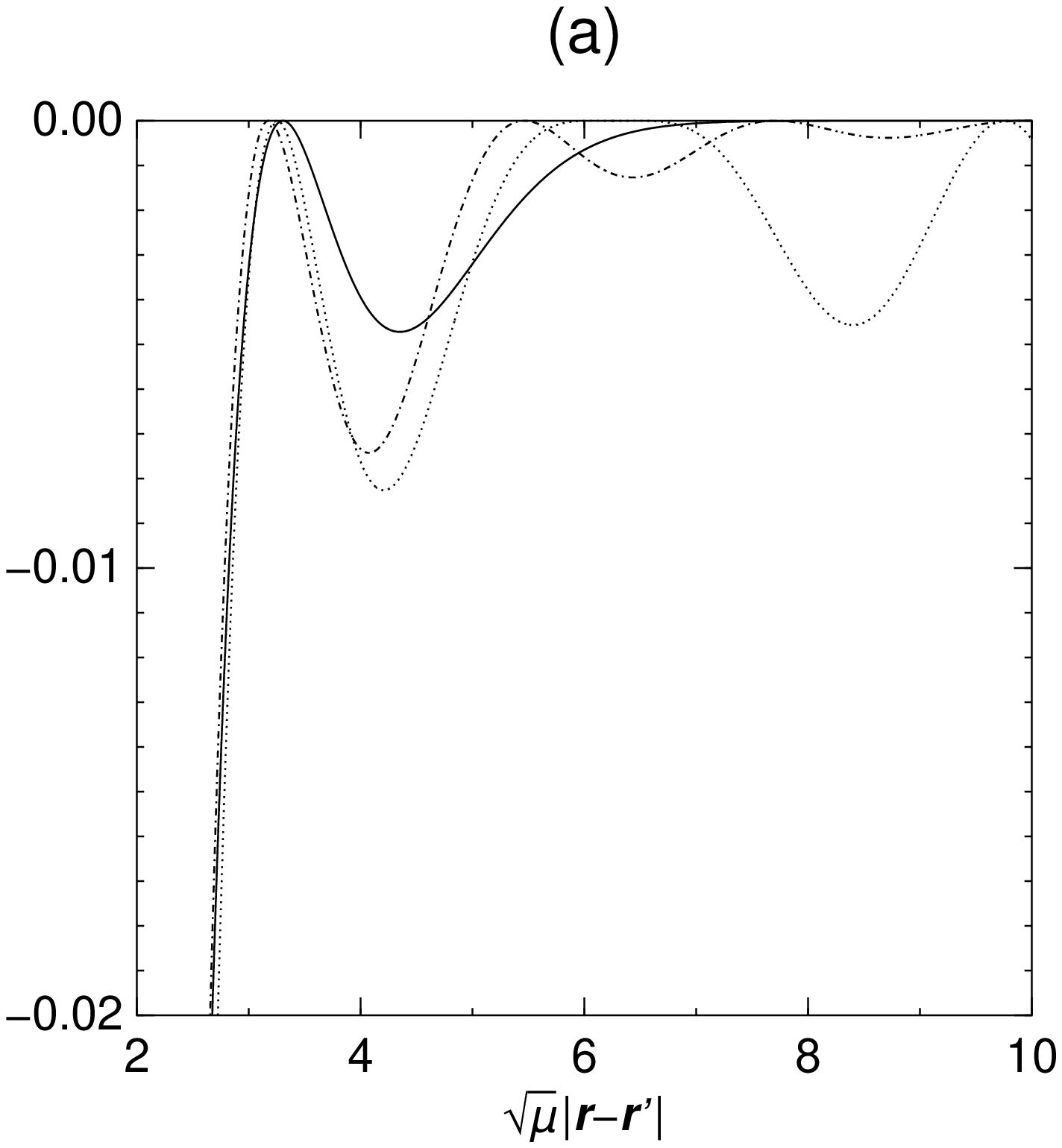}
    \hspace{0.5cm}
    \includegraphics[height=6.5cm]{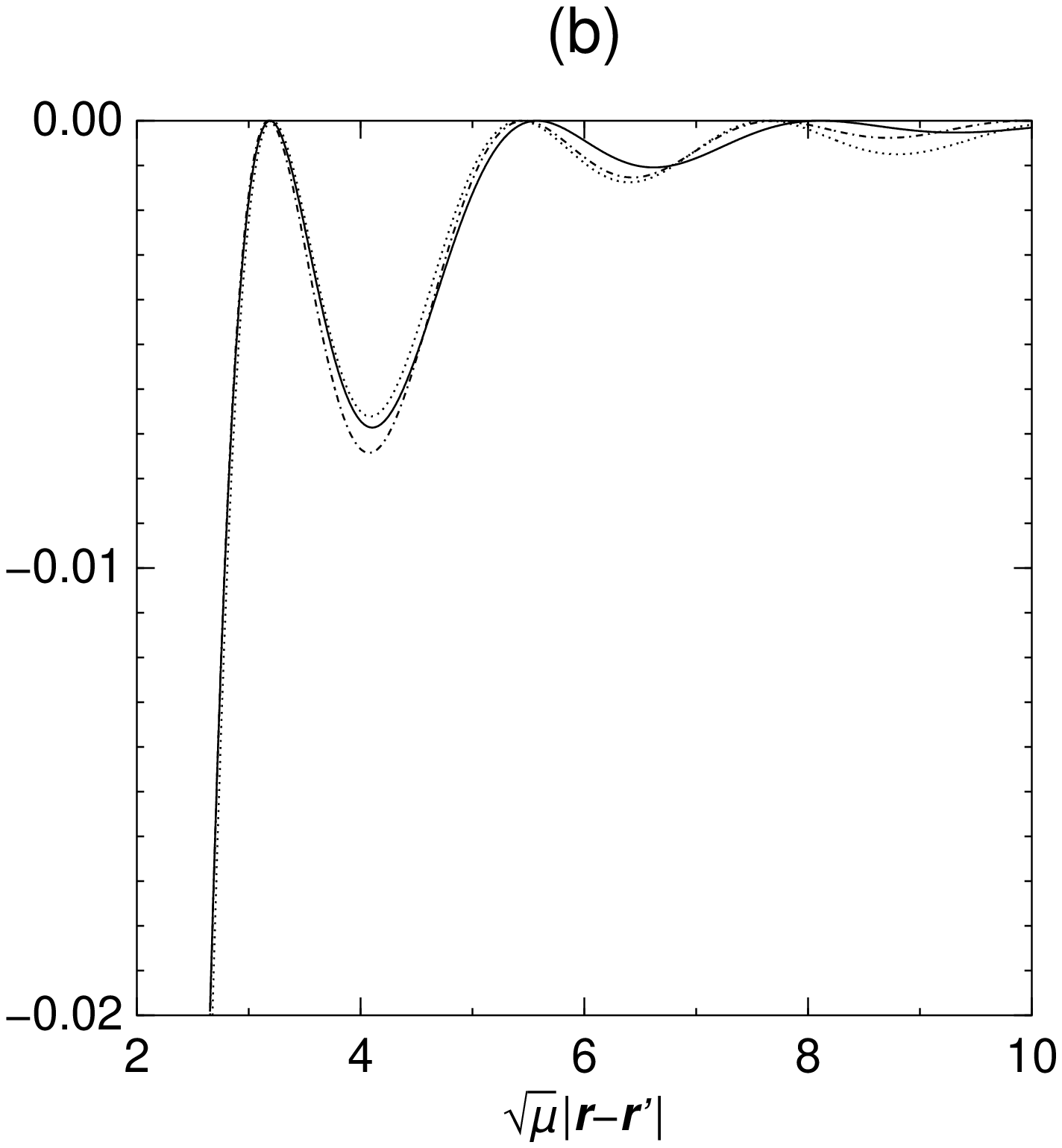}
  \end{center}
  \caption{Bulk correlation functions 
    $g(xyz, xy'z)$ (\full),
    $g(xyz, xyz')$ (\dotted) for $B\neq 0$ and (a) $\dlmu=2$, 
    (b) $\dlmu=5$, and $g(\vect{r}, \vect{r}')$ (\chain) for $B=0$.
    All curves start at $-1$ for $\vect{r} = \vect{r}'$.}
  \label{fig1}
\end{figure}

In figure~\ref{fig1} we have plotted the bulk correlation function for 
$B=0$ and for $B\neq0$ with $\nu=2$ and $\nu=5$. For non-vanishing magnetic 
field we focused on the correlation functions with position differences that 
are either parallel with or perpendicular to the magnetic field.
For large fields, or, more precisely, for small $\nu$, the correlation 
functions for the parallel and the perpendicular directions differ 
considerably. For somewhat larger $\nu$, however, the correlation functions 
become fairly similar, both in the nodal structure, and in the amplitudes. As 
it turns out, these similarities are manifest already for $\nu=5$, where the 
number of completely filled Landau levels is still rather low.

Comparing (\ref{3.14}) with (\ref{3.6}), we see that, by turning on the 
magnetic field, the range of the correlations in the plane perpendicular to 
the magnetic field becomes smaller, with a Gaussian instead of an algebraic 
decay. In contrast, the range of the correlations in the direction parallel 
to the magnetic field becomes somewhat larger. In fact, although the decay 
remains algebraic when the field is switched on, the dominant contribution 
in the tail of the correlation function becomes inversely proportional to 
the square of the distance, whereas it is inversely proportional to the 
fourth power of the distance in the field-free case.

\section{Path-decomposition expansion}
\label{sec:MRE}

Introducing the wall at $x=0$ makes the temperature Green function 
dependent on the distance from the wall, i.e.\ the coordinate $x$. In the 
absence of a magnetic field the influence of the wall on the Green function 
is easily found from a reflection principle \cite{ROEP:1996}. The 
temperature Green function becomes
\begin{equation}
\label{4.1}
  G_\beta(\vect{r},\vect{r}') =  G^b_\beta(\vect{r},\vect{r}') -
G^b_\beta(\vect{r},\vect{r}'')
\end{equation}
with the bulk Green function (\ref{3.15}) and the reflected position 
$\vect{r}''$ defined as $(x'',y'',z'')=(-x',y'z')$. Likewise, the 
$\mu$-dependent Green function gets the form
\begin{equation}
\label{4.2}
  G_\mu(\vect{r},\vect{r}') =  G^b_\mu(\vect{r},\vect{r}') -
G^b_\mu(\vect{r},\vect{r}'')
\end{equation}
with the bulk Green function (\ref{3.14}). 

When a magnetic field is present, the influence of the wall on the properties 
of the system is more difficult to determine. In \cite{KES:1999}, we have seen 
that the temperature Green function can be found from a path-decomposition 
expansion
\begin{equation}
  \label{4.3}
  G_\beta(\vect{r},\vect{r}') =
  \sum_{n=0}^\infty G^{(n)}_\beta(\vect{r}, \vect{r}')
\end{equation}
where $G^{(n)}_\beta(\vect{r}, \vect{r}')$ is the contribution from paths that 
hit the wall $n$ times. This path-decomposition expansion, which was first 
formulated in \cite{AUK:1985}, is fully equivalent to the multiple-reflection 
expansion as introduced in \cite{BAB:1970}, and discussed for a confined 
magnetic system in \cite{JOH:1995}. We have also seen in \cite{KES:1999} that 
for $\vect{r}$ and $\vect{r}'$ at large distances from the wall, the terms 
with small $n$ in (\ref{4.3}) are more important than those with larger $n$. 
In particular, the $n=1$ term will give us the leading-order correction on the 
bulk quantities. The latter correspond to $n=0$, so that one has  
$G^{(0)}_\beta(\vect{r}, \vect{r'}) = G^b_\beta(\vect{r}, \vect{r'})$. If no 
field is present the expansion terminates after the term with $n=1$, for all 
distances from the wall. 

In the particular case where $x=x'$, the transverse part of the $n=1$ term has 
the form \cite{JOH:1995}, \cite{KES:1999}
\begin{equation}
  \label{4.4}
\fl  G^{(1)}_{\perp,\beta}(xy, xy') =
  -\frac{B^2}{16\pi^{3/2}}
  \int_0^\beta \rmd\tau\,
  f^{(1)}_{\beta,\tau}(xy, xy')\,
  \exp\left[g^{(1)}_{\beta,\tau}(xy, xy')\right].
\end{equation}
The functions $f^{(1)}_{\beta,\tau}(xy, xy')$ and
$g^{(1)}_{\beta,\tau}(xy, xy')$ are given by
\begin{equation}
\label{4.5}
  f^{(1)}_{\beta, \tau} (xy, xy') =
  \frac{1}{2}\;
  \frac{(t_1 t_2)^{1/2}}{s_1 s_2 (t_1 + t_2)^{1/2}}\,
  \left[\left(\frac{1}{t_1} + \frac{1}{t_2}\right) \dlx + \rmi \dly\right]
\end{equation}
and
\begin{equation}
\label{4.6}
  g^{(1)}_{\beta,\tau}(xy, xy') =
  - \frac{1}{4}
  \left[
    \left(\frac{1}{t_1}+\frac{1}{t_2}\right) \dlx^2
    - 2\rmi \dlx\dly
    + \frac{\dly^2}{t_1+t_2}
  \right]
\end{equation}
with $t_1 = \tanh(\tau B/2)$, $s_1 = \sinh(\tau B/2)$, 
$t_2 =\tanh((\beta-\tau)B/2)$ and $s_2 = \sinh((\beta-\tau)B/2)$.

\subsection{Case $y \neq y'$}

If we set $p=q^2/({q_0}^2-q^2)$, with $q = \tanh[(2\tau-\beta)B/4]$ and $q_0 = 
\tanh(\beta B/4)$, we can write (\ref{4.4}) as
\begin{eqnarray}
\label{4.7}
  \fl G^{(1)}_{\perp,\beta}(xy, xy') =
  -\frac{B}{2^{7/2}\,\pi^{3/2}}\,
  \frac{1-{q_0}^2}{{q_0}^{3/2}}\,
  \exp\left(-\frac{\dlx^2}{2q_0}+\frac{\rmi \dlx\dly}{2}
   -\frac{1+{q_0}^2}{16q_0}\,\dly^2\right)
  \nonumber \\ \times
  \int_0^\infty \frac{\rmd p}{\sqrt{p(1+p)}}
  \left[
    \dlx \sqrt{1+(1-q_0^2)p}
    + \frac{\rmi\dly}{2}\, \frac{q_0}{\sqrt{1+(1-q_0^2)p}}
  \right]
  \nonumber \\ \times
  \exp\left[-\frac{(1-{q_0}^2)p}{2q_0}\,\dlx^2
  -\frac{1-{q_0}^2}{16q_0}\,\frac{1+(1+{q_0}^2)p}{1+(1-{q_0}^2)p}\,
    \dly^2\right].
\end{eqnarray}
When $\dlx$ is large, only small values of $(1-{q_0}^2)p$ contribute, which 
implies that we can set $\sqrt{1+(1-{q_0}^2)p} \approx 1$. Furthermore, we can 
replace the last exponential function with $\exp[-(1-{q_0}^2)\dly^2/(16q_0)]$, at least 
as long as $\dly$ is finite.  If we make those substitutions and use
\begin{equation}
\label{4.8}
  \int_0^\infty \frac{\rmd p}{\sqrt{p(1+p)}}\, \rme^{-a p} = 
\rme^{a/2} \, K_0(a/2)
\end{equation}
(where $K_0$ is the modified Bessel function of the second kind), we get
\begin{eqnarray}
  \label{4.9}
  G^{(1)}_{\perp,\beta}(xy, xy') \approx
  -\frac{B}{2^{7/2}\,\pi^{3/2}}
  \left(\dlx + \frac{\rmi\dly}{2} q_0\right)
  \frac{1-{q_0}^2}{{q_0}^{3/2}}
  \nonumber \\ \times
  \exp
  \left(
    -\frac{1+{q_0}^2}{4q_0} \dlx^2 +
    \rmi\frac{\dlx\dly}{2} -
    \frac{\dly^2}{8q_0}
  \right)
  K_0\left(\frac{1-{q_0}^2}{4q_0} \dlx^2\right).
\end{eqnarray}

If we are only interested in the non-degenerate case, where in general 
$(1-{q_0}^2)\dlx^2/q_0 $ is large, we can use the asymptotic expansion for 
$K_0$ \cite{MOS:1966}. Multiplying the result with the 
longitudinal Green function, which is the same as in the bulk, we find the 
first-order correction to the total temperature Green function in the 
approximate form
\begin{equation}
\label{4.10}
\fl  G^{(1)}_{\beta}(xyz,xy'z) \approx
  - \frac{B}{2^{7/2}\,\pi^{3/2}\,\beta^{1/2}} \,
  \frac{\sqrt{1-{q_0}^2}}{q_0}\,
  \exp
  \left(
    - \frac{\dlx^2}{2q_0}
    + \rmi\frac{\dlx\dly}{2}
    - \frac{\dly^2}{8 q_0}
  \right).
\end{equation}
However, for the degenerate case we need the full complexity of (\ref{4.9}).

In order to obtain results for the degenerate case, we now apply
(\ref{2.4}).  In doing so we choose the contour of integration
by setting $\beta = (\rmi t+1) \dlx / B$:
\begin{eqnarray}
\label{4.11}
  \fl G^{(1)}_\mu(xyz,xy'z) \approx
  - \frac{B^{3/2}}{16\pi^3\,\dlx^{1/2}} \rme^{\rmi\dlx\dly/2}
  \int_{-\infty}^\infty \rmd t\,
  \frac{\rme^{\dlmu\dlx(\rmi t + 1)}}{(\rmi t + 1)^{3/2}}
  \left(
    \dlx + \frac{\rmi\dly}{2}q_0
  \right)
  \frac{1-{q_0}^2}{{q_0}^{3/2}}
  \nonumber \\ \times
  \exp\left(-\frac{1+{q_0}^2}{4q_0} \dlx^2 -\frac{\dly^2}{8q_0}\right)
  K_0\left(\frac{1-{q_0}^2}{4q_0}\dlx^2\right).
\end{eqnarray}
In the new variables $q_0$ equals $\tanh[(\rmi t+1)\dlx/4]$.  From $\dlx \gg 
1$ one finds $q_0 \approx 1$, which in turn implies $1-q_0^2 \approx 
4\exp[-\dlx(\rmi t+1)/2]$.  This also means that for finite $\dly$ we may 
replace $\dlx + \rmi \dly q_0/2$ with $\dlx$. With the help of the series 
representation of the modified Bessel function
\begin{equation}
  \label{4.12}
  K_0(u) = \sum_{n=0}^\infty \left[\sum_{m=1}^n \frac{1}{m} - \gamma -
    \log \left(\frac{u}{2}\right) \right] \frac{1}{2^{2n}(n!)^2}\, u^{2n}
\end{equation}
and the integral relation
\begin{equation}
  \label{4.13}
  \int_{-\infty}^\infty \rmd t \, \frac{\rme^{(\rmi t+1)x}}{(\rmi t+1)^\nu} = 
\frac{2\pi\, x^{\nu-1}}{\Gamma(\nu)}\, \theta(x)   \qquad (\nu > 0)
\end{equation}
we arrive at the asymptotic expression
\begin{eqnarray}
  \label{4.14}
  \fl G^{(1)}_{\mu}(xyz, xy'z) \approx
  -\frac{B^{3/2}\,\dlx}{\pi^{5/2}}
  \rme^{-\dlx^2/2}\,
  \rme^{\rmi\dlx\dly/2}\,
  \rme^{-\dly^2/8}\,
  {\sum_n}'\; \frac{\dlx^{4n}}{2^{2n}(n!)^2}
  \nonumber \\ \times
  \sqrt{\nu-(n+1/2)} \left[ \frac{1}{4[\dlmu-(n+1/2)]} + \sum_{m=1}^n
    \frac{1}{m} - \gamma - \ln\left(\frac{\dlx^2}{2}\right)\right]
\end{eqnarray}
which is valid for large $\sqrt{B} x$. Again we recognize the sum over 
Landau levels.  In fact, apart from the phase $\rme^{\rmi \dlx\dly/2}$ and the 
factor $\rme^{-\dly^2/8}$, the result is identical to what we found in 
\cite{KES:1999} for $\vect{r}=\vect{r}'$.

\subsection{Case $z \neq z'$}

This case is considerably simpler than the $y \neq y'$-case, since the 
temperature Green function factorizes.  The transverse part of the Green 
functions follows from (\ref{4.9}) by taking $\dly = 0$. Using the asymptotic  
expansion for the Bessel function and introducing the longitudinal part of the 
Green function, one finds the total temperature Green function for the 
non-degenerate case with $z\neq z'$ as
\begin{equation}
\label{4.15}
  \fl G^{(1)}_\beta(xyz,xyz') \approx
  -\frac{B}{2^{7/2}\,\pi^{3/2}\,\beta^{1/2}}\,
  \frac{\sqrt{1-{q_0}^2}}{q_0}
  \exp\left(-\frac{\dlx^2}{2q_0} - \frac{\dlz^2}{2\beta B}\right)
\end{equation}
which is the analogue of (\ref{4.10}). 

To calculate the $\mu$-dependent Green function $G^{(1)}_\mu(xyz, xyz')$ for 
the degenerate case we need to keep the Bessel function in (\ref{4.9}).  With 
$s=\beta B$ we get
\begin{eqnarray}
\label{4.16}
  \fl G^{(1)}_\mu(xyz, xyz') \approx
  - \frac{B^{3/2}\,\dlx}{8\pi^2}\,
  \frac{1}{2\pi\rmi} \int_{c-\rmi\infty}^{c+\rmi\infty} \rmd s\,
  \frac{\rme^{\dlmu s}}{s^{3/2}}\,
  \frac{1-{q_0}^2}{{q_0}^{3/2}}\,
  \exp\left(-\frac{1+{q_0}^2}{4q_0}\dlx^2\right)
  \nonumber \\ \times
  K_0\left(\frac{1-{q_0}^2}{4q_0}\dlx^2\right)\,
  \exp\left(-\frac{\dlz^2}{2 s}\right).
\end{eqnarray}
We are free to choose the contour of integration, as long as it is in the 
right half-plane.  By choosing $c$ large (i.e. $c = \dlx$), we can make the 
same approximations for $q_0$ and $1-{q_0}^2$ as before.  In terms of the 
integration variable $s$ these approximations read $q_0 \approx 1$ and 
$1-{q_0}^2 \approx 4 \rme^{-s/2}$.  Together with (\ref{4.12}) this yields
\begin{eqnarray}
\label{4.17}
  \fl G^{(1)}_\mu(xyz,xyz') \approx
  -\frac{B^{3/2}\,\dlx}{2\pi^2}\,
  \rme^{-\dlx^2/2}
  \sum_{n=0}^\infty\, \frac{\dlx^{4n}}{2^{2n}(n!)^2}\,
  \frac{1}{2\pi\rmi} \int_{\dlx-\rmi\infty}^{\dlx+\rmi\infty} \rmd s\,
  \rme^{[\dlmu-(n+1/2)]s}
  \nonumber \\ \times
  \left\{
    \frac{1}{2} \frac{1}{s^{1/2}} +
    \left[ \sum_{m=1}^n \frac{1}{m} - \gamma - \ln\left(\frac{\dlx^2}{2}\right)\right]
    \frac{1}{s^{3/2}}
  \right\}\,
  \exp\left(-\frac{\dlz^2}{2s}\right).
\end{eqnarray}
The integral in this expression is still an inverse Laplace transform.  With 
the help of (\ref{3.5}) and the analogous identity \cite{ERD:1954}
\begin{equation}
\label{4.18}
  \frac{1}{2\pi\rmi} \int_{c-\rmi\infty}^{c+\rmi\infty} \rmd s\, 
\rme^{st}\,
  s^{-1/2}\, \rme^{-a/s} =
  \frac{1}{\sqrt{\pi t}}\, \cos\big(2\sqrt{at}\, \big)\, \theta(t) 
\end{equation}
for $a > 0$ (and $c>0$) we finally get the asymptotic expression
\begin{eqnarray}
  \label{4.19}
  \fl G^{(1)}_\mu(xyz,xyz') \approx
-\frac{B^{3/2}\,\dlx}{\pi^{5/2}}\, \rme^{-\dlx^2/2}
  {\sum_n}'\; \frac{\dlx^{4n}}{2^{2n}(n!)^2} \,\sqrt{\nu-(n+1/2)}\nonumber\\
\times  \left\{ \frac{\cos\big(\sqrt{2[\nu-(n+1/2)]}\,\zeta\big)}{4[\nu-(n+1/2)]}
  \right.
  \nonumber \\ +
  \left.
\frac{\sin\big(\sqrt{2[\nu-(n+1/2)]}\,\zeta\big)}{\sqrt{2[\nu-(n+1/2)]}\,\zeta}\;
  \left[ \sum_{m=1}^n \frac{1}{m} - \gamma - \ln\left(\frac{\dlx^2}{2}\right)\right]
  \right\}.
\end{eqnarray}
for large $\sqrt{B} x$. This result looks a bit more complicated than 
(\ref{4.14}). In the limit $\dlz \rightarrow 0$ we recover $G^{(1)}_\mu(xyz,
xyz)$, as calculated in \cite{KES:1999}. 

Before we discuss the expressions (\ref{4.14}) and (\ref{4.19}) in more 
detail, we will first show that the same results can be derived using a 
different approach.

\section{Parabolic cylinder functions}

Results equivalent to those of the previous section can be derived by using 
the explicit representation of the temperature Green function in terms of 
parabolic cylinder functions, which are the eigenfunctions of the transverse 
part of the system Hamiltonian. Because of translation invariance in the 
$y$-direction it is convenient to use a Fourier transform and to write the 
transverse part of the Hamiltonian as
\begin{equation}
\label{5.1}
  H_\perp(k) = - \frac{1}{2} \frac{\partial^2}{\partial x^2}
  + \frac{1}{2} (Bx - k)^2.
\end{equation}
For fixed $k$, this Hamiltonian with boundary condition $\psi(k,x=0)=0$ 
defines an eigenvalue problem. In terms of the eigenfunctions $\psi_n(k,x)$ 
and the corresponding eigenvalues $E_n(k)$ we define the 
Fourier-transformed transverse energy Green function \cite{KES:1998}
\begin{equation}
\label{5.2}
  G_{\perp,E}(k,x,x') = \sum_n \psi_n(k, x) \psi_n^*(k, x')\;
\delta\left[E_n(k) - E\right]
\end{equation}
where the eigenfunctions are normalized to $\int_0^{\infty}dx\, 
\abs{\psi_n(k,x)}^2=1$. In terms of this energy Green function the (total) 
temperature Green function (\ref{2.1}) is
\begin{eqnarray}
\label{5.3}
  \fl G_\beta(\vect{r}, \vect{r}') = (2\pi\beta)^{-1/2}\,
  \exp[-(z-z')^2/2\beta] \nonumber\\
\times\int_0^\infty \rmd E \, \rme^{-\beta E}\,
  \frac{1}{2\pi} \int_{-\infty}^{\infty} \rmd k\, \rme^{\rmi k(y-y')}\,
  G_{\perp,E}(k,x,x').
\end{eqnarray}
Performing the inverse Laplace transform as in (\ref{2.4}) and using 
(\ref{3.5}), we obtain
\begin{equation}
  \label{5.4}
  \fl G_\mu(\vect{r}, \vect{r}') =
  \frac{1}{\pi^2}
  \int_0^{\mu} \rmd E\,
  \frac{\sin\big[\sqrt{2(\mu - E)}(z-z')\big]}{z-z'} \,
  \int_{-\infty}^{\infty} \rmd k\, \rme^{\rmi k (y-y')} \,
  G_{\perp,E}(k, x, x').
\end{equation}
We now choose $x=x'$, as before, and switch again to the dimensionless 
variables $\dlx$, $\dlz$ (and a dimensionless Fourier variable 
$\dlk=k/\sqrt{B}$ as well). In terms of these, the energy Green function is 
\cite{KES:1998}
\begin{equation}
\label{5.5}
  \fl G_{\perp,E} (k,x,x) =  \sqrt{B} \, \sum_n \,
\frac{D^2_{\dlE_n(k)-1/2}\left[\sqrt{2}(\dlx-\dlk)\right]}
{\int_0^\infty \rmd \dlx'\,
    D^2_{\dlE_n(\dlk)-1/2}\left[\sqrt{2}(\dlx'-\dlk)\right]}\; 
  \delta\left[E - B \dlE_n(\dlk)\right].
\end{equation}
Here $D_{\lambda}(u)$ is a parabolic cylinder function \cite{MOS:1966}. 
Furthermore, $\dlE_n(\dlk)$ is determined by the boundary condition
\begin{equation}
  \label{5.6}
  D_{\dlE_n(\dlk)-1/2}(-\sqrt{2} \dlk) = 0.
\end{equation}
If we now carry out the integration over $E$ in (\ref{5.4}), we get
\begin{eqnarray}
  \label{5.7}
  \fl G_\mu(xyz,xy'z') =
  \frac{B^{3/2}}{\pi^2 }\;
  {\sum_n}'
  \int_{\dlk_n(\dlmu)}^\infty \rmd \dlk\,
  \rme^{\rmi \dlk\dly}\; 
  \frac{\sin\big(\sqrt{2[\dlmu-\dlE_n(\dlk)]}\,\dlz\big)}{\dlz}\nonumber\\
\times  \frac{D^2_{\dlE_n(k)-1/2}\left[\sqrt{2}(\dlx-\dlk)\right]}
{\int_0^\infty \rmd \dlx'\,
    D^2_{\dlE_n(\dlk)-1/2}\left[\sqrt{2}(\dlx'-\dlk)\right]}
\end{eqnarray}
where $\dlk_n(\dlmu)$ is determined by 
$\dlE_n\left[\dlk_n(\dlmu)\right]=\dlmu$.
 
\begin{figure}
  \begin{center}
    \includegraphics[height=6.5cm]{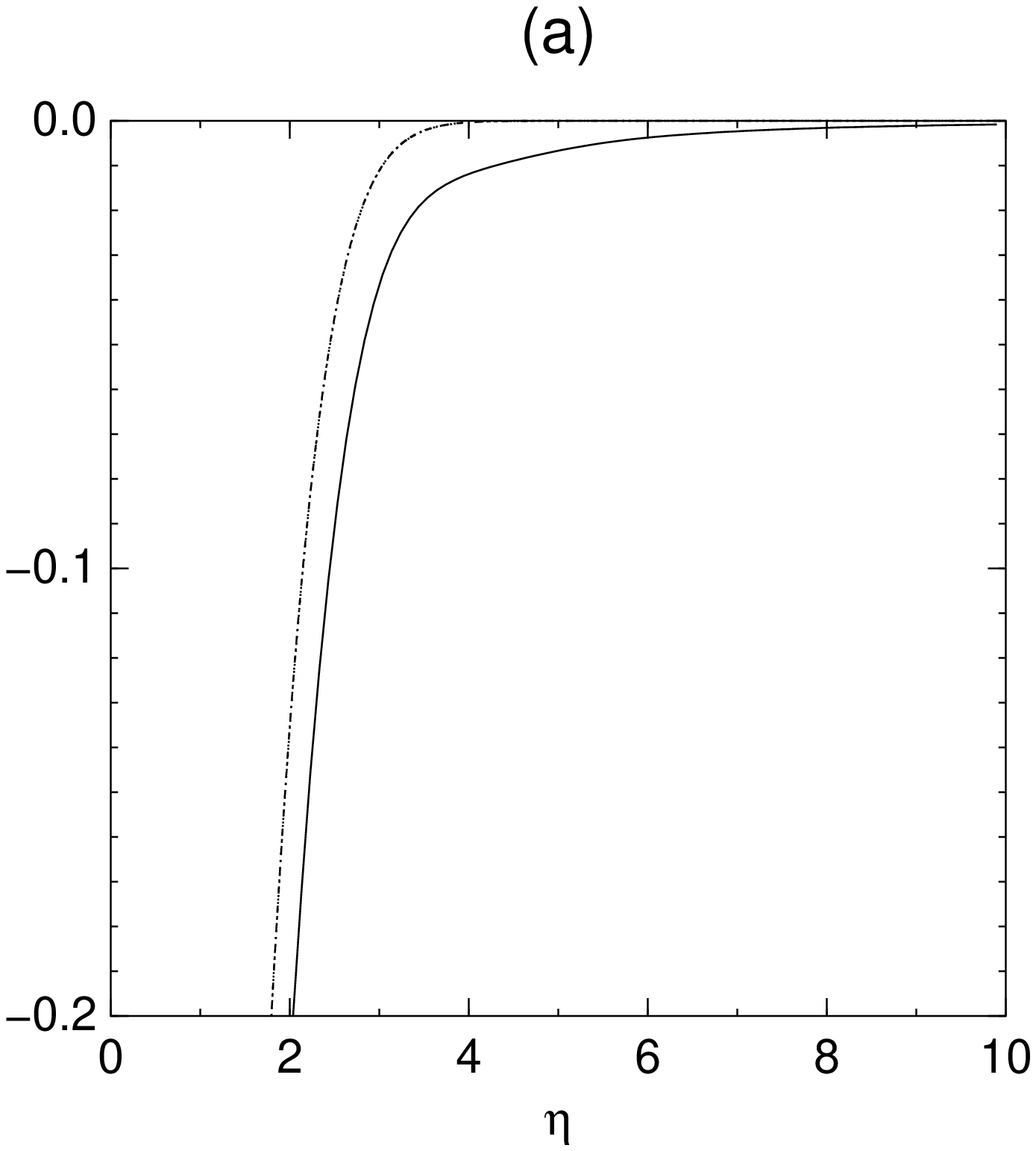}
    \hspace{0.5cm}
    \includegraphics[height=6.5cm]{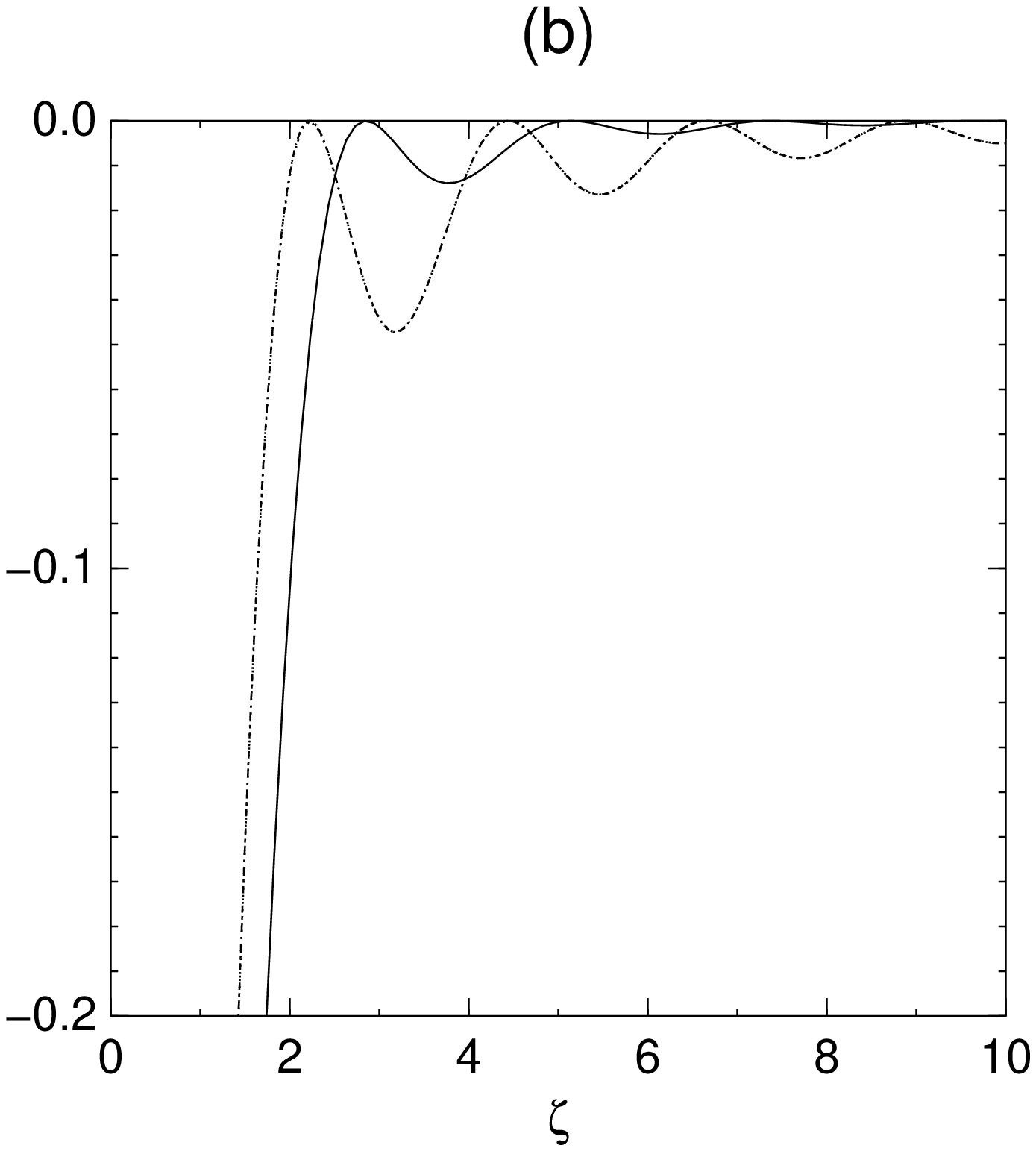}
  \end{center}
  \caption{Correlation functions (a) $g(xyz,xy'z)$ and (b) $g(xyz,xyz')$ for 
$\dlx=1$ (\full), $\dlx=5$ (\chain) and in the bulk ($\dlx\rightarrow\infty$,
    \dashed) for $\dlmu=1.5$.  The curves for $\dlx=5$ and for the bulk
    are (almost) indistinguishable.}
  \label{fig2}
\end{figure}

In figure~\ref{fig2} we have plotted the resulting correlation function, 
which we calculated numerically for several values of $\dlx$ and for a value 
of the chemical potential, which  corresponds to a completely filled lowest 
Landau level.
Even at moderately small distances from the wall the influence of the 
confinement on the correlation function is not very big.  In the $y$-direction 
the correlations become a little stronger. The same is true for the 
correlations in the $z$-direction, at least for small $\dlz$. However, 
the oscillating tail in the correlation function is suppressed by the presence 
of the wall.

\subsection{Case $y \neq y'$}

Taking the limit $z' \rightarrow z$ or $\dlz \rightarrow 0$ in (\ref{5.7}) is 
trivial. The resulting formula contains a phase-factor $\rme^{\rmi \dlk\dly}$, 
which is absent in the case $y=y'$ as considered in \cite{KES:1998}.  We first 
split off the bulk contribution
\begin{equation}
\label{5.8}
  \fl G^b_\mu(xyz,xy'z) =
  \frac{\sqrt{2}B^{3/2}}{\pi^2}\;
  {\sum_n}'
  \int_{-\infty}^\infty \rmd \dlk\,
  \rme^{\rmi (\dlk+\dlx)\dly}
  \frac{\sqrt{\nu-(n+1/2)}}{\sqrt{\pi}\, n!}\,
  D^2_n(-\sqrt{2}\dlk).
\end{equation}
The remainder $G^c_\mu(xyz,xy'z) = G_\mu(xyz,xy'z) - G^b_\mu(xyz,xy'z)$ gives 
the correction due to the wall.

In order to obtain an asymptotic expansion of $G^c_\mu(xyz,xy'z)$ for large 
$x$ we split the integration at $\dlk'=\alpha' \dlx$, with $0 < \alpha' < 
1-\frac{1}{2}\sqrt{2}$, and $\dlk''= \alpha''\dlx$, with $\alpha'' > 
\frac{1}{2}\sqrt{2}$.  As in the case with $\vect{r}=\vect{r}'$, the 
contributions from the intervals $[\dlk_n(\dlmu),\dlk']$ and $[\dlk'',
\infty)$ decay faster than $\exp(-\dlx^2/2)$. With the help of the asymptotic 
expressions \cite{KES:1998}
\begin{equation}
  \label{5.9}
\fl  \dlE_n(\dlk) - (n+\case{1}{2}) \approx
  \frac{1}{\sqrt{\pi}\, n!}\, 2^n \,\rme^{-\dlk^2} \,\dlk^{2n+1}
\end{equation}
\begin{eqnarray}
  \label{5.10}
  \fl \left\{
    \int_0^\infty \rmd \dlx'\,
    D^2_{\dlE_n(\dlk)-1/2}[\sqrt{2}(\dlx'-\dlk)]
  \right\}^{-1} \nonumber\\
\approx  \frac{1}{\sqrt{\pi}\, n!} -
  \frac{1}{\pi (n!)^2}\, 2^{n+1}\, \rme^{-\dlk^2}\, \dlk^{2n+1}\,
  \left[
    \sum_{m=1}^{n} \frac{1}{m} - \gamma - \ln(\sqrt{2}\dlk)
  \right]
\end{eqnarray}
and
\begin{eqnarray}
\label{5.11}
  \fl D^2_{\dlE_n(\dlk)-1/2}[\sqrt{2}(\dlx-\dlk)] \approx
  2^n\, \rme^{-(\dlx-\dlk)^2}\, (\dlx-\dlk)^{2n}\nonumber\\
  + \frac{1}{\sqrt{\pi}\, n!}\, 2^{2n+1} \,\rme^{-(\dlx-\dlk)^2}\, 
\rme^{-\dlk^2}\,
  (\dlx-\dlk)^{2n}\, \dlk^{2n+1}\, \ln\big[\sqrt{2}(\dlx-\dlk)\big]
\end{eqnarray}
one finds
\begin{eqnarray}
  \label{5.12}
  \fl G^c_\mu(xyz,xy'z) \approx 
  \frac{\sqrt{2}B^{3/2}}{\pi^2}\;
  {\sum_n}' \; \frac{2^{2n+1}}{\pi(n!)^2}\,
  \sqrt{\dlmu-(n+1/2)}
  \nonumber \\ \times
  \int_{\dlk'}^{\dlk''} \rmd \dlk\,
  \rme^{\rmi \dlk\dly}\,
  \rme^{-\dlk^2}\,
  \rme^{-(\dlx-\dlk)^2}\,
  \dlk^{2n+1} \,(\dlx-\dlk)^{2n}\,
  P_n(\dlk,\dlx-\dlk)
\end{eqnarray}
with
\begin{equation}
  \label{5.13}
  \fl P_n(\dlk,\dlx-\dlk) =
  - \left\{ \frac{1}{4[\mu-(n+1/2)]}
    + \sum_{m=1}^n \frac{1}{m}
    - \gamma
    - \ln \big[2\dlk(\dlx-\dlk)\big]
  \right\}.
\end{equation}
We now expand the integrand in (\ref{5.12}) around $\dlk=\dlx/2$:
\begin{eqnarray}
\label{5.14}
\fl  \int_{\dlk'}^{\dlk''} \rmd \dlk\,
  \rme^{\rmi \dlk \dly}\, \rme^{-\dlk^2}\,
  \rme^{-(\dlx-\dlk)^2}\,
  \dlk^{2n+1}\, (\dlx - \dlk)^{2n}\,
  P_n(\dlk, \dlx-\dlk) 
  \nonumber \\
  \approx  \rme^{\rmi \dlx\dly/2}\,
  \rme^{-\dlx^2/2}\,
  (\dlx/2)^{4n+1}\,
  P_n(\dlx/2, \dlx/2)
  \int_{-\infty}^\infty \rmd t\, \rme^{\rmi t\dly - 2 t^2}
\end{eqnarray}
since the main contribution comes from the region around $\dlk=\dlx/2$, at 
least for $\dlx \gg 1$.  Evaluating the remaining integral gives us exactly 
(\ref{4.14}). 

\begin{figure}
  \begin{center}
    \includegraphics[height=6.5cm]{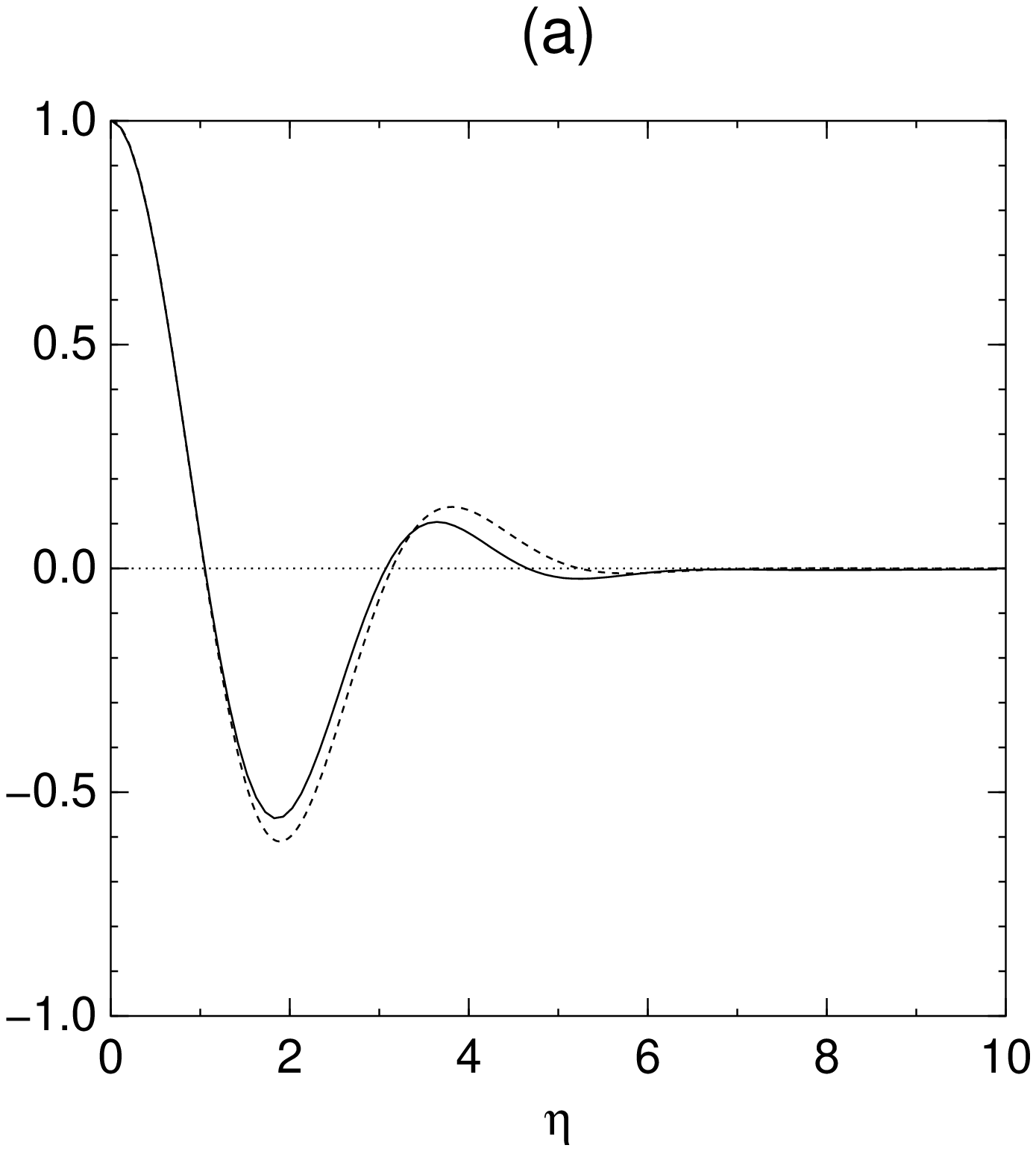}
    \hspace{0.5cm}
    \includegraphics[height=6.5cm]{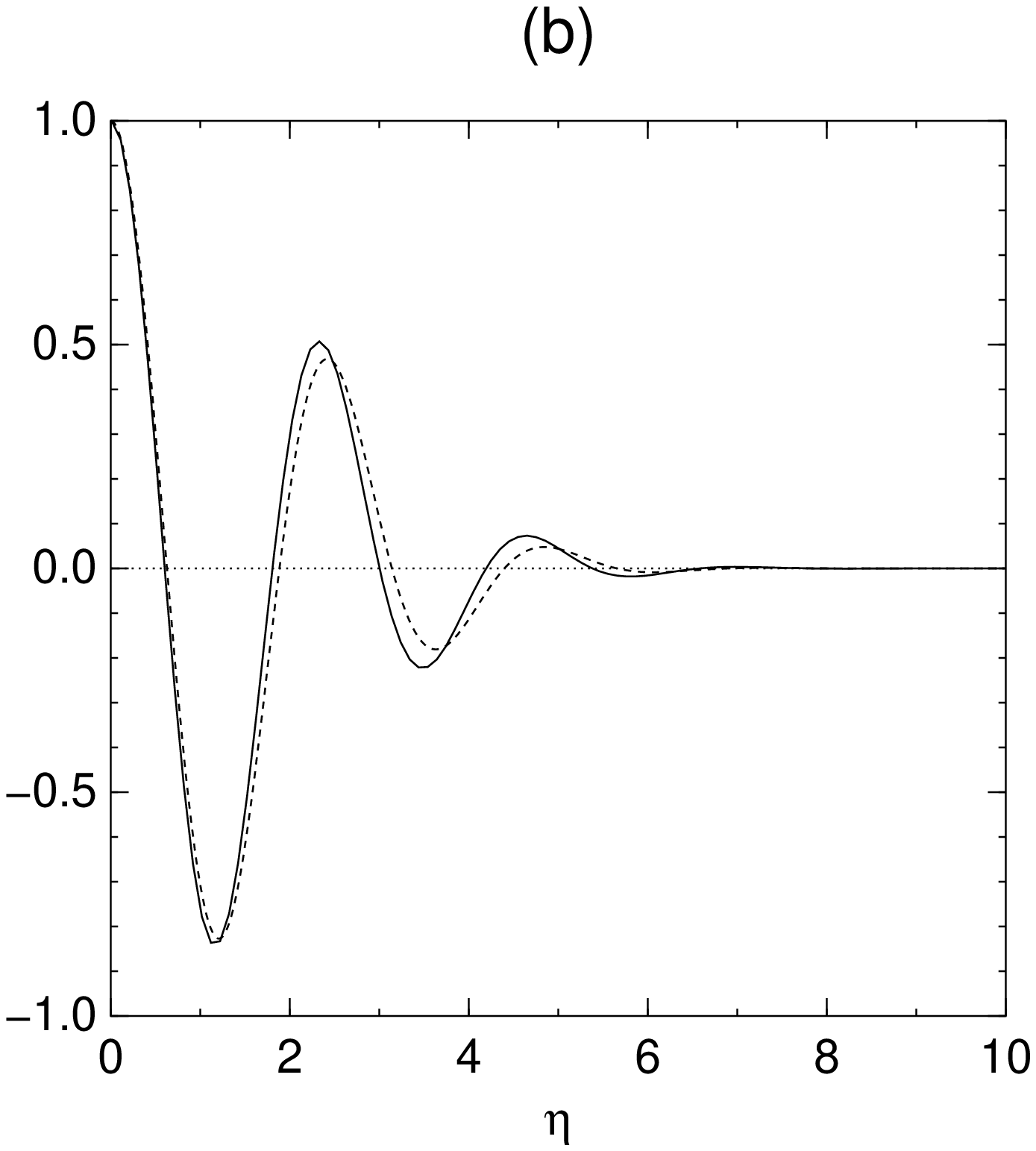}
  \end{center}
  \caption{Comparison between numerical data (\full) and asymptotic
    form (\dashed) of the real part of $G^c_\mu(xyz,xy'z)/G^c_\mu(xyz,xyz)$ at 
(a)$\dlx=3$ and (b) $\dlx=5$ for $\dlmu=1.5$.}
\label{fig3}
\end{figure}

Note that in this case, instead of $G^{(1)}_\mu(xyz,xy'z)$, we have in fact 
calculated $G^c_\mu(xyz,xy'z)$.  However since the terms beyond $n=1$ in the 
path-decomposition expansion (\ref{4.3}) are of higher order, the asymptotic 
form of $G^c_\mu(xyz,xy'z)$ is in leading order identical to that of 
$G^{(1)}_\mu(xyz,xy'z)$. However, in the present case we can easily take along 
more terms in (\ref{5.13}) (see \cite{KES:1998}). The expansion around 
$\dlk=\dlx/2$ given here is only valid for $\dly \ll \dlx$.  If $\dly$ is of 
the same order of magnitude as $\dlx$, the terms that we left out would not be 
small compared to the leading term and we would not get the right asymptotics. 
In figure \ref{fig3} we compare the results of a numerical evaluation of the 
exact expression for the real part of the wall correction $G^c_\mu(xyz,
xy'z)/G^c_\mu(xyz,xyz)$, which follows from (\ref{5.7}) and (\ref{5.8}), and 
the asymptotic form in leading order, as derived in (\ref{4.14}). As expected, 
the agreement is best for small $\dly$. The imaginary part of $G^c_\mu(xyz,
xy'z)/G^c_\mu(xyz,xyz)$ behaves in a similar way. 

\begin{figure}
  \begin{center}
    \includegraphics[height=6.5cm]{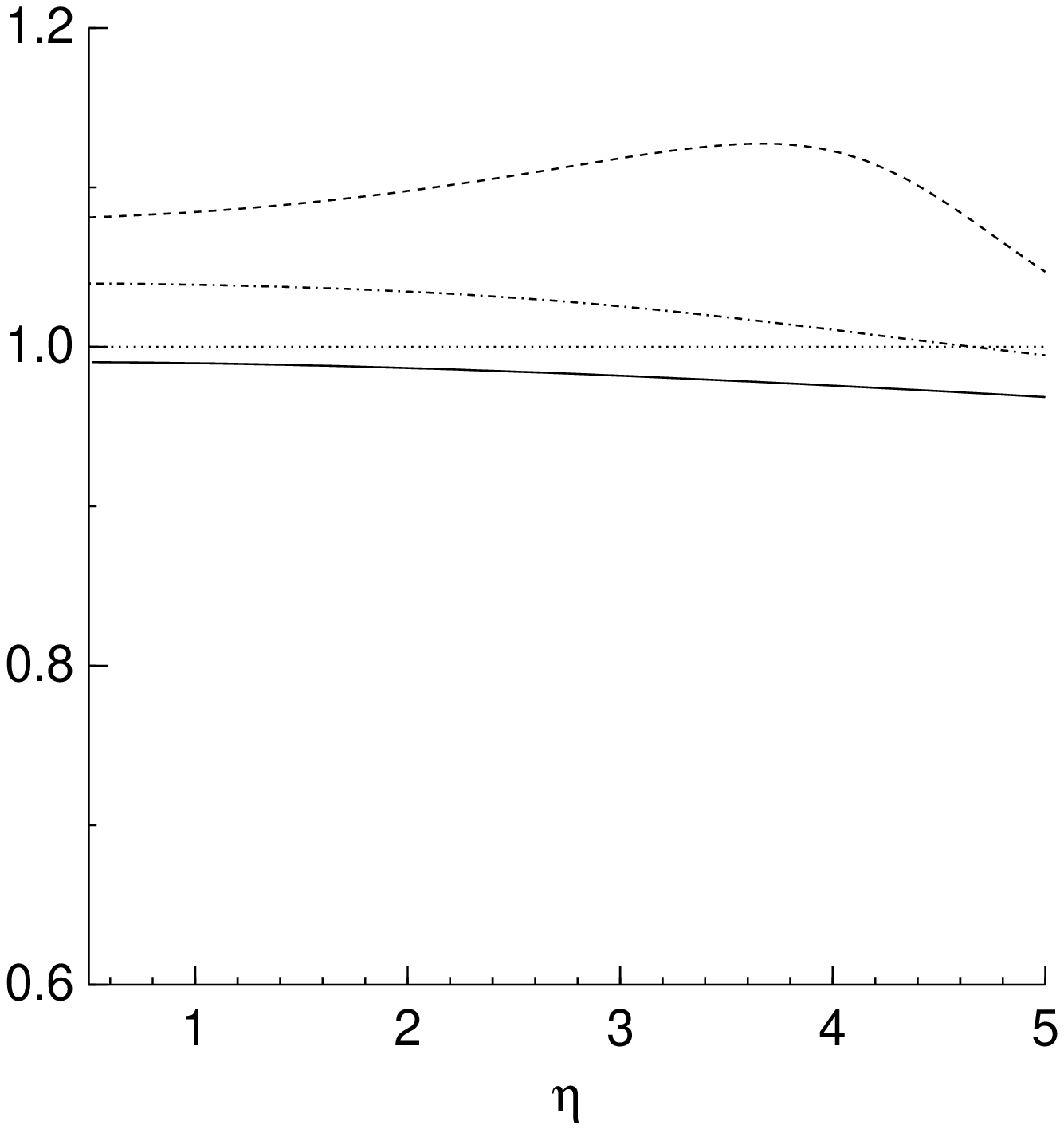}
  \end{center}
  \caption{Check of the Gaussian decay, by evaluation of 
$\sqrt{8}\dly^{-1}[\log(\abs{G^c_\mu(xyz,xy'z)} /\abs{G^c_\mu(xyz,
xyz)})]^{1/2}$, for $\dlmu=1.5$ and $\dlx=5$ (\full), $\dlx=4$ (\chain), 
$\dlx=3$ (\dashed). The dotted line (\dotted) is the asymptotic value according 
to (\ref{4.14}).}
  \label{fig4}
\end{figure}

A striking feature of the asymptotic expansion (\ref{4.14}) is its Gaussian  
decay proportional to $\rme^{-\dly^2/8}$. It is slower than the decay of the 
bulk $\mu$-dependent Green function (\ref{3.6}), which is proportional to 
$\rme^{-\dly^2/4}$. This somewhat slower decay is indeed corroborated by the 
numerical evaluation of $G^c_\mu(xyz,xy'z)/G^c_\mu(xyz,xyz)$, as is shown in 
figure \ref{fig4}. For small $\dly$ and large $\dlx$ (i.e.\ for the regime 
where (\ref{4.14}) holds) the curves converge to the asymptotic value 
$1$.

\subsection{Case $z \neq z'$}

On setting $y=y'$ in (\ref{5.7}) the phase-factor $\rme^{\rmi k\dly}$ drops 
out.  With the help of the same asymptotic expressions as in the case with $y 
\neq y'$, and the same splitting of the integration interval, we arrive at the 
asymptotic expression
\begin{eqnarray}
\label{5.15}
  \fl G^c_\mu(xyz,xy'z) \approx 
  \frac{\sqrt{2}B^{3/2}}{\pi^2}
  {\sum_n}' \frac{2^{2n+1}}{\pi(n!)^2}\;
  \sqrt{\dlmu-(n+1/2)}
  \nonumber \\ \times
  \int_{\dlk'}^{\dlk''} \rmd \dlk\,
  \rme^{-\dlk^2} \,
  \rme^{-(\dlx-\dlk)^2} \,
  \dlk^{2n+1}\,(\dlx-\dlk)^{2n}\,
  Q_n(\dlk,\dlx-\dlk, \dlz)
\end{eqnarray}
with
\begin{eqnarray}
\label{5.16}
  \fl Q_n(\dlk,\dlx-\dlk,\dlz) =
  - \frac{\cos\big(\sqrt{2[\dlmu-(n+1/2)]}\,\dlz \big)}
  {4[\dlmu-(n+1/2)]}
  - \frac{\sin\big(\sqrt{2[\dlmu-(n+1/2)]}\,\dlz \big)}
  {\sqrt{2[\dlmu-(n+1/2)]}\, \dlz}
  \nonumber \\ \times
  \left\{
     \sum_{m=1}^n \frac{1}{m} - \gamma - \ln \big[2\dlk(\dlx-\dlk)\big]
  \right\}.
\end{eqnarray}
Again, the main contribution to the integral comes from the region around 
$\dlk=\dlx/2$.  Expansion around this point allows us to evaluate the 
integral, and we recover (\ref{4.19}).

\begin{figure}
  \begin{center}
    \includegraphics[height=6.5cm]{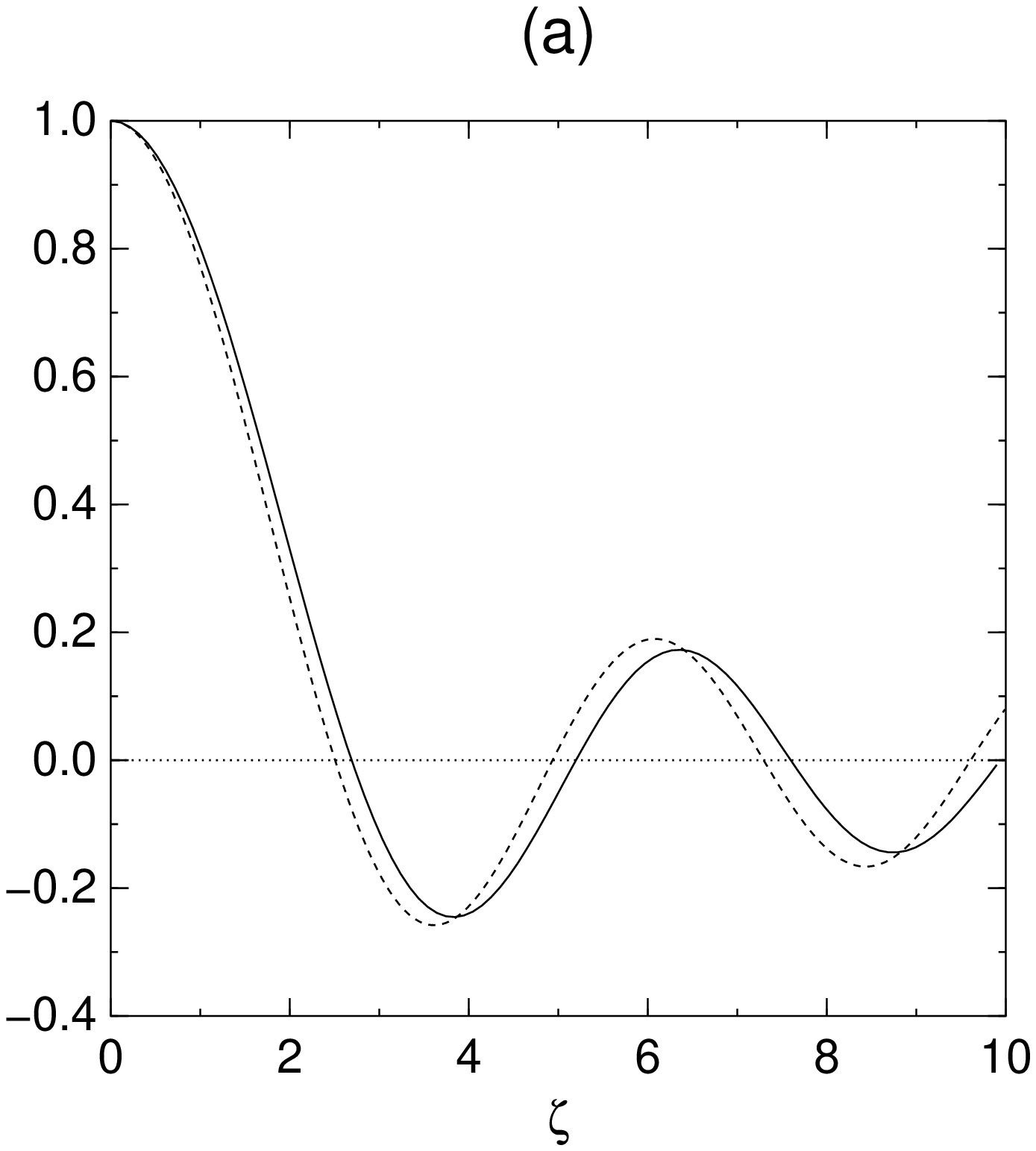}
    \hspace{0.5cm}
    \includegraphics[height=6.5cm]{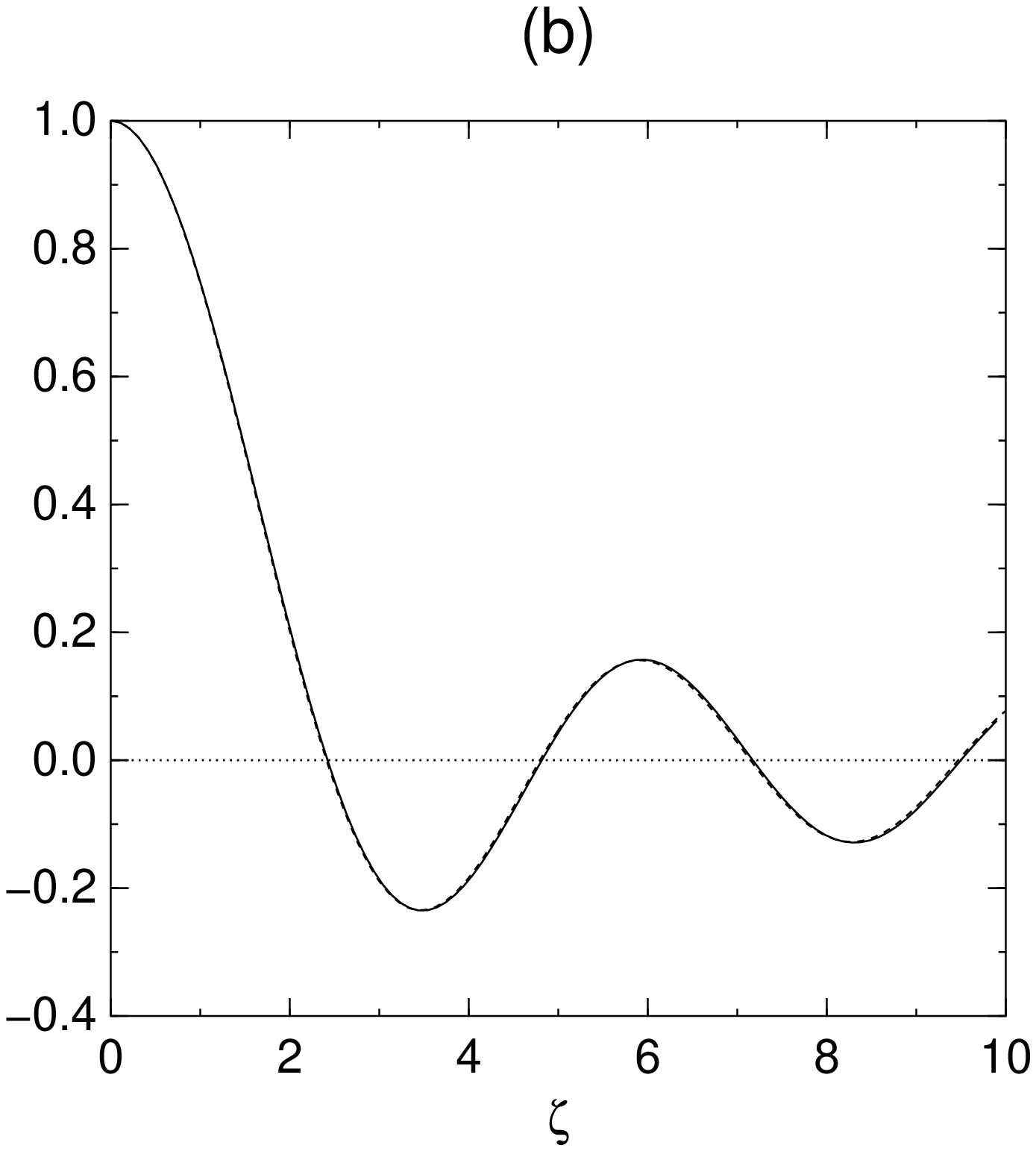}
  \end{center}
  \caption{Comparison between numerical data (\full) and asymptotic
    form (\dashed) of $G^c_\mu(xyz,xyz')/G^c_\mu(xyz,xyz)$ at (a)
    $\dlx=3$ and (b) $\dlx=5$ for $\dlmu=1.5$. In the latter case the 
    two curves are (almost) indistinguishable.}
  \label{fig5}
\end{figure}

In figure~\ref{fig5} we compare the asymptotic expression (\ref{4.19}) 
for $G^c_\mu(xyz,xyz')$ with numerical data.
As expected, the differences get smaller for increasing values of $\dlx$.

\section{Correlations for large separations $\abs{\vect{r}-\vect{r}'}$}

The path-decomposition expansion, which we employed in section \ref{sec:MRE}, 
is expedient for large $x$ only. However, the range of validity of the 
representation of the previous section is not limited to that regime, so that 
we can use it for small $x$ as well. In particular, it is helpful in 
determining the asymptotic behaviour of the correlation function for finite 
$x$ and large distances between the points of observation, both in the $y$- 
and the $z$-direction. 

\subsection{Large $\abs{y-y'}$}
Let us go back to (\ref{5.7}). In order to bring the square-root singularity 
at $\dlk=\dlk_n(\dlmu)$ to the fore, we write
\begin{equation}
  \label{6.1}
  G_\mu(xyz,xy'z) =
  \frac{\sqrt{2} B^{3/2}}{\pi^2} {\sum_n}'
  \int_{\dlk_n(\dlmu)}^\infty \rmd \dlk\, \rme^{\rmi \dlk \dly}
  \sqrt{\dlk-\dlk_n(\dlmu)}\, \phi_n(\dlk)
\end{equation}
with
\begin{equation}
\label{6.2}
  \phi_n(\dlk) = 
\frac{\sqrt{\dlmu-\dlE_n(\dlk)}}{\sqrt{\dlk-\dlk_n(\dlmu)}}\; 
  \frac{D^2_{\dlE_n(\dlk)-1/2}\big[\sqrt{2}(\dlx-\dlk)\big]}
  {\int_0^\infty \rmd \dlx'\,
    D^2_{\dlE_n(\dlk)-1/2}\big[\sqrt{2}(\dlx'-\dlk)\big]}.
\end{equation}
The function $\phi_n(\dlk)$ is analytic on the integration interval, and 
proportional to $\dlk^{2n-1/2} \rme^{-\dlk^2}$ for $\dlk \rightarrow\infty$.  
Hence, the asymptotics of (\ref{6.1}) for large $\abs{\dly}$ are determined by 
the lower boundary of the integral only. With the help of the method of 
stationary phase \cite{ERD:1956} we find
\begin{eqnarray}
  \label{6.3}
  \fl \int_{\dlk_n(\dlmu)}^\infty \rmd \dlk\, \rme^{\rmi \dlk \dly} \,
  \sqrt{\dlk-\dlk_n(\dlmu)}\, \phi_n(\dlk) \approx
  {\abs{\dly}}^{-3/2}\,
  \rme^{\rmi [\dlk_n(\dlmu) \dly + (3\pi/4)\sgn(\dly)]} \,
  \Gamma(\case{3}{2})
  \nonumber \\ \times
  \left.
    \sqrt{-\frac{\rmd \dlE_n(\dlk)}{\rmd \dlk}}
  \right|_{\dlk=\dlk_n(\dlmu)}
  \frac{D^2_{\dlmu-1/2}\big[\sqrt{2}(\dlx-\dlk_n(\dlmu))\big]}
  {\int_0^\infty \rmd \dlx'\, 
D^2_{\dlmu-1/2}\big[\sqrt{2}(\dlx'-\dlk_n(\dlmu))\big]}
\end{eqnarray}
where we used that the derivative of $\dlE_n(\dlk)$ is negative. In 
\cite{KES:1998} we have derived the identity
\begin{eqnarray}
\label{6.4}
  \fl \left\{
    \int_0^\infty \rmd \dlx'\,
    D^2_{\dlE_n(\dlk)-1/2}(\sqrt{2}(\dlx'-\dlk))
  \right\}^{-1} \nonumber\\
= - \frac{1}{2\pi}\, \Gamma^2\big[-\dlE_n(\dlk)+\case{1}{2}\big]\,
  D^2_{\dlE_n(\dlk)-1/2}\big(\sqrt{2}\dlk\big)\,
    \frac{\rmd \dlE_n(\dlk)}{\rmd \dlk}.
\end{eqnarray}
This equality allows us to get rid of the normalization integral in 
(\ref{6.3}). In this way we arrive at the following asymptotic expression 
for the $\mu$-dependent Green function at large $\abs{\dly}$ and finite 
$\dlx$:
\begin{eqnarray}
\label{6.5}
  \fl G_\mu(xyz,xy'z) \approx 
  \frac{B^{3/2}\, \rme^{\rmi 
(3\pi/4)\sgn(\dly)}}{2^{3/2}\,\pi^{5/2}}\,
\Gamma^2\big(-\dlmu+\case{1}{2}\big)\;\frac{1}{{\abs{\dly}}^{3/2}}\;
{\sum_n}' \rme^{\rmi \dlk_n(\dlmu) \dly}\nonumber\\
\times D^2_{\dlmu-1/2}\big[\sqrt{2}\dlk_n(\dlmu)\big]\,
D^2_{\dlmu-1/2}\big[\sqrt{2}(\dlx-\dlk_n(\dlmu))\big]\,
\left[-\frac{d \dlk_n(\dlmu)}{d\dlmu}\right]^{-3/2}.
\end{eqnarray}

\begin{figure}
  \begin{center}
    \includegraphics[height=6.5cm]{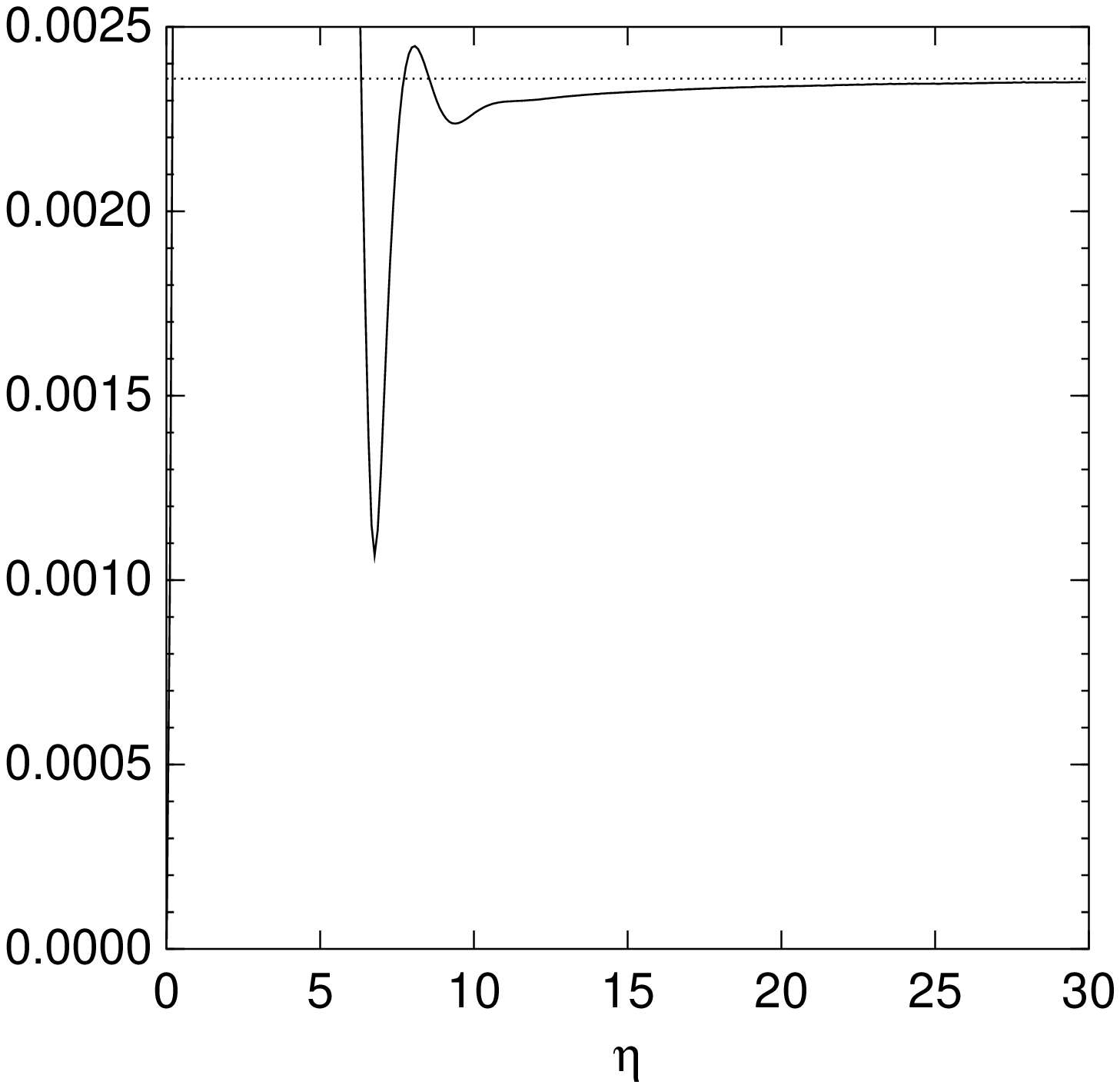}
  \end{center}
  \caption{Comparison of $\dly^{3/2} \pi^2 \abs{G^c_\mu(xyz,xy'z)} / (\sqrt{2}
    B^{3/2})$ (\full) at $\dlx=3$, $\dlmu=1.5$  and its asymptotic value 
(\dotted) for large $\abs{\dly}$, according to (\ref{6.6})}
  \label{fig6}
\end{figure}

The asymptotic expression simplifies considerably for the special case of a 
completely filled lowest Landau level, that is for $\dlmu=3/2$ and $n=0$. By 
using the representation of the parabolic cylinder function in terms of 
confluent hypergeometric functions \cite{MOS:1966} we find $\dlk_0(\dlmu)=0$ 
and $d\dlk_0(\dlmu)/d\dlmu=-\sqrt{\pi}/2$. Employing (\ref{6.4}) and inserting 
$D_1(\sqrt{2}u)=\sqrt{2}\, u\, \exp(-u^2/2)$, we  arrive at the simple 
asymptotic expression
\begin{equation}  
\label{6.6}
G_\mu(xyz,xy'z) \approx \frac{4 B^{3/2}\, 
\rme^{\rmi (3\pi/4)\sgn(\dly)}} {\pi^{9/4}}\, \dlx^2 \, \rme^{-\dlx^2} \, 
\frac{1}{\abs{\dly}^{3/2}}.
\end{equation}
for large $\abs{\dly}$ and $\dlmu=3/2$. The asymptotic behaviour 
proportional to ${\abs{\dly}}^{-3/2}$ in (\ref{6.5}) and (\ref{6.6}) is 
clearly induced by the presence of the wall, since the decay of the bulk 
$\mu$-dependent Green function in the $y$-direction is Gaussian, at least 
in the presence of a magnetic field (see (\ref{3.6})). The algebraic decay 
is corroborated by a numerical evaluation of (\ref{5.7}). Both results are 
compared in figure~\ref{fig6}.

\subsection{Large $\abs{z-z'}$}

To determine the asymptotics for large separations in the $z$-direction we set 
$y=y'$ in (\ref{5.7}). Subsequently, we need to determine the asymptotic 
behaviour of the imaginary part of the integral
\begin{equation}
\label{6.7}
  I = \int_{\dlk_n(\dlmu)}^{\infty} \rmd \dlk\; \rme^{\rmi
    \sqrt{2\left[\dlmu-\dlE_n(\dlk)\right]}\, \abs{\dlz}}\;
  \frac{D^2_{\dlE_n(\dlk)-1/2}\big[\sqrt{2}(\dlx-\dlk)\big]}{\int_0^\infty \rmd
    \dlx'\, D^2_{\dlE_n(\dlk)-1/2}\big[\sqrt{2}(\dlx'-\dlk)\big]}.
\end{equation}
for large $\abs{\dlz}$. In order to obtain the asymptotics, we split the 
integration interval at $\dlk' \gg 1$.  For $\dlk \geq \dlk'$ we can use 
(\ref{5.9}) and (\ref{5.10}). In the numerator we insert the asymptotic 
expression for the parabolic cylinder function
\begin{equation}
\label{6.8}
  \fl D_{\dlE_n(\dlk)-1/2}(\sqrt{2}(\dlx-\dlk)) \approx 2^{n/2+1}(-1)^n \dlk^n
  \rme^{-\dlk^2/2} \sinh\big(\dlk\dlx-\dlx^2/2\big),
\end{equation}
which is valid for large $\dlk$ and finite $\dlx$. This relation can be 
derived by using (\ref{5.9}), and the asymptotic relations 
\begin{equation}
\label{6.9}
  D_n(\sqrt{2}u) \approx (-1)^n\, 2^{n/2}\, \abs{u}^n \,\rme^{-u^2/2}
\end{equation}
and
\begin{equation}
\label{6.10}
  \left. \frac{\partial D_\nu(\sqrt{2}u)}{\partial \nu} \right|_{\nu=n} 
\approx
  (-1)^{n+1}\, 2^{-n/2}\, \sqrt{\pi}\, n!\, \abs{u}^{-n-1}\, \rme^{u^2/2}
\end{equation}
which are both valid for large negative $u$. In this way the contribution from 
$\dlk \geq \dlk'$ to (\ref{6.7}) becomes
\begin{eqnarray}
\label{6.11}
  \fl I_1 \approx
  \int_{\dlk'}^\infty \rmd \dlk \,
  \exp\left(\rmi\sqrt{2[\dlmu-(n+1/2)]}\,\abs{\dlz} -
  \rmi\frac{2^{n-1/2}\,\rme^{-\dlk^2} \,\dlk^{2n+1}}
  {\sqrt{\pi}\, n!\,\sqrt{\dlmu-(n+1/2)}}\; \abs{\dlz}
  \right)
  \nonumber \\ \times
  \frac{2^{n+2}}{\sqrt{\pi}\,n!}\;
  \dlk^{2n}\, \rme^{-\dlk^2}\, \sinh^2\big(\dlk\dlx-\dlx^2/2\big).
\end{eqnarray}
We now introduce a new  integration variable $t=\rme^{-\dlk^2}\, \dlk^{2n+1}$, 
which implies  $\dlk \approx\sqrt{-\ln t}$ for large $\dlk$. The integral then 
gets the form 
\begin{eqnarray}
\label{6.12}
  \fl I_1 \approx \frac{2^{n+1}}{\sqrt{\pi}\,n!}\;
  \rme^{\rmi \sqrt{2[\dlmu-(n+1/2)]}\,\abs{\dlz}}\;
  \int_0^{t'} \rmd t\,
  \exp\left(
    -\rmi \frac{2^{n-1/2}\abs{\dlz}}{\sqrt{\pi}\, n!\,
      \sqrt{\dlmu-(n+1/2)}}\; t
  \right)
  \nonumber \\ \times
  \frac{\sinh^2\big(\sqrt{-\ln t}\, \dlx - \dlx^2/2\big)}{- \ln t}
\end{eqnarray}
with $t'=\rme^{-{\dlk'}^2}\, {\dlk'}^{2n+1} \ll 1$.  Note that there is
a logarithmic singularity at $t=0$.

The asymptotic expansion of an integral of the form
\begin{equation}
\label{6.13}
  \int_0^{t'} \rmd t\,
  \rme^{\rmi u t}
  (-\ln t)^\mu
\end{equation}
for $t' < 1$ and large $u$ has a contribution from the lower boundary and 
the upper boundary. The contribution from the lower boundary is 
\cite{WON:1989}
\begin{equation}
  \label{6.14}
  \frac{\rmi}{u} \sum_{r=0}^\infty (-1)^r \binom{\mu}{r}
  \left[
    \sum_{k=0}^r \binom{r}{k} \Gamma^{(k)}(1)
    \left(\frac{\pi\rmi}{2}\right)^{r-k}
  \right]
  (\ln u)^{\mu - r}
\end{equation}
where $\Gamma^{(k)}(u)$ is the $k$-th derivative of $\Gamma(u)$. For functions 
$f(-\ln t)$ that can be expressed as a Laurent series in $\sqrt{-\ln t}$ for 
small $t$, one obtains the asymptotics of the integral
\begin{equation}
\label{6.15}
  \int_0^{t'} \rmd t\, \rme^{\rmi u t}\, f(-\ln t)
\end{equation}
for large $u$ by integrating the series term-by-term by means of 
(\ref{6.14}). In fact, the contribution from the lower boundary of the 
integral is
\begin{equation}
\label{6.16}
  \frac{\rmi}{u} \sum_{r=0}^\infty \frac{(-1)^r}{r!}
  \left[
    \sum_{k=0}^r \binom{r}{k} \Gamma^{(k)}(1)
    \left(\frac{\pi\rmi}{2}\right)^{r-k}
  \right]
  \frac{\rmd^r f(\ln u)}{\rmd(\ln u)^r}.
\end{equation}

Let us now return to (\ref{6.12}), which contains an integral that has the 
form of the complex conjugate of (\ref{6.15}). For this particular case one 
has $f(-\ln t) = \sinh^2(\sqrt{-\ln t}\,\dlx -\dlx^2/2)/(- \ln t)$ and $u= 
2^{n-1/2} \abs{\dlz}/[\sqrt{\pi}\, n!\, \sqrt{\dlmu-(n+1/2)}]$. The 
term with $r=0$ in (\ref{6.16}) dominates the contribution from the lower 
boundary to the asymptotics, since $f(\ln u)$ is much bigger than all its 
derivatives when $u$ is large. Using this fact we get for the contribution 
from the lower boundary ($t=0$) to the asymptotics of (\ref{6.12})
\begin{eqnarray}
\label{6.17}
  \fl -\rmi\,\frac{2^{n+1}}{\sqrt{\pi}\, n!\, u}
  \rme^{\rmi \sqrt{2[\dlmu-(n+1/2)]}\, \abs{\dlz}}\;
  \frac{\sinh^2\big(\sqrt{\ln u}\,\dlx - \dlx^2/2\big)}{\ln u} 
  \nonumber \\
\rule{-1cm}{0cm}  \approx-\rmi \,
  \frac{2 \sqrt{2[\dlmu-(n+1/2)]}}{\abs{\dlz}\;\ln\abs{\dlz}}\,
  \rme^{\rmi \sqrt{2[\dlmu-(n+1/2)]}\, \abs{\dlz}}\,
  \sinh^2\big(\sqrt{\ln\abs{\dlz}}\, \dlx - \dlx^2/2\big).
\end{eqnarray}
The contribution from the upper boundary ($t=t'$) to the asymptotics of 
(\ref{6.12}) can be found by use of the same theorem from \cite{WON:1989}. It 
is expected to cancel against the contribution from the end-point $\dlk=\dlk'$ 
of the integral
\begin{equation}
\label{6.18}
  I_2 = \int_{\dlk_n(\dlmu)}^{\dlk'} \rmd \dlk\; \rme^{\rmi
    \sqrt{2\left[\dlmu-\dlE_n(\dlk)\right]}\, \abs{\dlz}}\;
  \frac{D^2_{\dlE_n(\dlk)-1/2}\big[\sqrt{2}(\dlx-\dlk)\big]}{\int_0^\infty \rmd
    \dlx'\, D^2_{\dlE_n(\dlk)-1/2}\big[\sqrt{2}(\dlx'-\dlk)\big]}.
\end{equation}
Indeed, it is not too difficult to show that it does, by evaluating the 
contribution from the end-point at $\dlk=\dlk'$ in $I_2$ with the help of 
standard techniques \cite{ERD:1956}. Finally, we have to check whether the 
end-point at $\dlk = \dlk_n(\dlmu)$ contributes to the asymptotics of $I_2$. 
The phase has a square-root singularity at that point. Using that fact, one 
finds that all terms in the asymptotic expansion that originate from the lower 
boundary of $I_2$ are real. Hence, they drop out when taking the imaginary 
part of $I$. 

Collecting our results, we have derived the asymptotic equality
\begin{equation}
\label{6.19}
  \fl \Imag I \approx
  -\frac{2 \sqrt{2[\dlmu-(n+1/2)]}}{\abs{\dlz}\;\ln\abs{\dlz}}\,
  \cos\big(\sqrt{2[\dlmu-(n+1/2)]}\, \dlz\big)\,
  \sinh^2\big(\sqrt{\ln\abs{\dlz}}\,\dlx - \dlx^2/2\big)
\end{equation}
which is valid for large $\abs{\dlz}$. As a consequence we obtain the 
$\mu$-dependent Green function
\begin{eqnarray}
\label{6.20}
  \fl G_\mu(xyz,xyz') \approx
  -\frac{2B^{3/2}}{\pi^2}\,
  \frac{\sinh^2\big(\sqrt{\ln{\abs{\dlz}}}\,\dlx-\dlx^2/2\big)}
  {\dlz^2\, \ln\abs{\dlz}}\nonumber\\
\times  {\sum_{n}}' \;\sqrt{2[\dlmu-(n+1/2)]}\,
  \cos\big(\sqrt{2[\dlmu-(n+1/2)]}\,\dlz\big).
\end{eqnarray}
for large $\abs{\dlz}$. This asymptotic form is valid at large separations 
$\abs{z-z'}$ and a finite distance $x$ from the wall. A  comparison with 
(\ref{3.6}) shows that the asymptotic behaviour changes substantially owing to 
the presence of the wall. The simple algebraic tail (modulated by a 
goniometric factor) which is valid in the bulk, is replaced by a more subtle 
decay involving a logarithmic dependence on $\abs{z-z'}$. 

\section{Correlations near the wall}

Let us return again to (\ref{5.7}). Using (\ref{6.4}) we may write it as
\begin{eqnarray}
  \label{7.1}
  \fl G_\mu(xyz,xy'z') =
  -\frac{B^{3/2}}{2\pi^3 }
  {\sum_n}'
  \int_{\dlk_n(\dlmu)}^\infty \rmd \dlk\,
  \rme^{\rmi \dlk\dly}\, 
  \frac{\sin\big(\sqrt{2[\dlmu-\dlE_n(\dlk)]}\,\dlz\big)}{\dlz}\,
  \Gamma^2\big[-\dlE_n(\dlk)+\case{1}{2}\big]
\nonumber\\
\times 
  D^2_{\dlE_n(\dlk)-1/2}\big(\sqrt{2}\dlk\big)\;
    D^2_{\dlE_n(\dlk)-1/2}\big[\sqrt{2}(\dlx-\dlk)\big] \,
    \frac{\rmd \dlE_n(\dlk)}{\rmd \dlk}.
\end{eqnarray}
For small $\dlx$ the last parabolic cylinder function can be expanded in a 
Taylor series. According to (\ref{5.6}) the zeroth-order term of this series 
vanishes, so that the leading term for small $\dlx$ is of first order in 
$\dlx$. The derivative of the parabolic cylinder function occurring in this 
term fulfills the Wronskian identity \cite{MOS:1966}
\begin{equation}
\label{7.2}
\fl\left. D_{\dlE_n(\dlk)-1/2}\big(\sqrt{2}\dlk\big) \, 
\frac{\partial}{\partial \dlk}
D_\nu\big(-\sqrt{2}\dlk\big) \right|_{\nu=\dlE_n(\dlk)-1/2}
=\frac{2\sqrt{\pi}}{\Gamma\big[-\dlE_n(\dlk)+\case{1}{2}\big]}
\end{equation}
where (\ref{5.6}) has been used again. Solving for the derivative we obtain
\begin{equation} 
  \label{7.3}
  \fl G_\mu(xyz,xy'z') \approx
  - \frac{2B^{3/2}\,\dlx^2}{\pi^2 }
  {\sum_n}'
  \int_{\dlk_n(\dlmu)}^\infty \rmd \dlk\,
  \rme^{\rmi \dlk\dly}\; 
  \frac{\sin\big(\sqrt{2[\dlmu-\dlE_n(\dlk)]}\, \dlz\big)}{\dlz} \, 
  \frac{\rmd \dlE_n(\dlk)}{\rmd \dlk}.
\end{equation}
The right-hand side can be simplified further by introducing the integration 
variable $\dlE$ instead of $\dlk$. In this way we arrive at the following 
approximate form of the $\mu$-dependent Green function near the wall:
\begin{equation} 
  \label{7.4}
  \fl G_\mu(xyz,xy'z') \approx
  \frac{2B^{3/2}\,\dlx^2}{\pi^2 }
  {\sum_n}'
  \int_{n+1/2}^{\dlmu} \rmd \dlE\,
  \rme^{\rmi \dlk_n(\dlE)\dly}\; 
  \frac{\sin\big[\sqrt{2(\dlmu-\dlE)}\,\dlz\big]}{\dlz}.
\end{equation}
As before we consider separately the cases $\dly\neq 0$ and $\dlz\neq 0$. 

\subsection{Case $y\neq y'$}
For $z'\rightarrow z$ the Green function becomes
\begin{equation} 
  \label{7.5}
  \fl G_\mu(xyz,xy'z) \approx
  \frac{2^{3/2}\,B^{3/2}\,\dlx^2}{\pi^2 }
  {\sum_n}'
  \int_{n+1/2}^{\dlmu} \rmd \dlE\, 
  \rme^{\rmi \dlk_n(\dlE)\dly}\; \sqrt{\dlmu-\dlE}.
\end{equation}
For large $\abs{\dly}$ the asymptotic behavior of the integral is determined by the 
value of the integrand near the upper boundary. One finds
\begin{equation}
  \label{7.6}
  \fl G_\mu(xyz,xy'z) \approx
  \frac{\sqrt{2}B^{3/2}\,\rme^{\rmi (3\pi/4)\sgn(\dly)}\; 
  \dlx^2}{\pi^{3/2} {\abs{\dly}}^{3/2}}\;
  {\sum_n}'\; \rme^{\rmi \dlk_n(\dlmu)\dly}\; 
\left[-\frac{d\dlk_n(\dlmu)}{d\dlmu}\right]^{-3/2}.
\end{equation}
The same result can be obtained from (\ref{6.5}) by expanding the last parabolic 
cylinder function for small $\dlx$ and using (\ref{7.2}).

The asymptotic decay proportional to ${\abs{\dly}}^{-3/2}$ in (\ref{7.6}) 
seems to be at variance with the behaviour of the correlations in the 
field-free case. For the latter case one finds
\begin{equation}
\label{7.7}
G_\mu(xyz,xy'z) \approx
-\frac{4\mu \, x^2}{\pi^2 (y-y')^3}\, \sin\big[\sqrt{2\mu}\,(y-y')\big].
\end{equation}
 by expanding (\ref{4.2}) with (\ref{3.14}) for small $x$ and retaining the 
terms dominant at large $\abs{y-y'}$. Hence, the correlations near the wall 
decay faster in the free-field case than in the case with field. It should be 
noted that (\ref{7.6}) is valid for $\abs{\dly}\gg 1$ or $\abs{y-y'}\gg 
B^{-1/2}$. For $B$ tending to 0 the region of validity thus shifts towards 
$\infty$. Furthermore, the number of terms in the sum becomes very large for 
small $B$ (since $\dlmu$ gets large), whereas the factor in front tends to 0, 
so that taking the limit $B\rightarrow 0$ in (\ref{7.6}) is not trivial. 

The decay of the field-free correlations can be derived in the present context 
by starting from (\ref{7.5}). Let us interchange the summation and the 
integral and split the latter at $\dlE'\gg 1$. We get
\begin{eqnarray}
\label{7.8}
\fl {\sum_n}'
  \int_{n+1/2}^{\dlmu} \rmd \dlE\, 
  \rme^{\rmi \dlk_n(\dlE)\dly}\; \sqrt{\dlmu-\dlE}=
\int_{1/2}^{\dlE'}\rmd \dlE\,  \sqrt{\dlmu-\dlE} \sum_{n=0}^{[\dlE-1/2]}\; 
\rme^{\rmi \dlk_n(\dlE)\dly}  \nonumber\\
+\int_{\dlE'}^{\dlmu}\rmd \dlE\,  \sqrt{\dlmu-\dlE}\sum_{n=0}^{[\dlE-1/2]}\; 
\rme^{\rmi \dlk_n(\dlE)\dly}.
\end{eqnarray}
In the second part at the right-hand side the sum consists of many terms for 
each $\dlE$. As discussed in the appendix, the values of $\dlk_n(\dlE)$ are 
located in the interval $[-\sqrt{2\dlE},\sqrt{2\dlE}]$, at least for all 
$n\leq[\dlE-1/2]-1$. As a consequence, their average separation goes to zero 
proportional to $1/\sqrt{\dlE}$. It is therefore expedient to replace the 
summation over $n$ by an integration over a continuous variable 
$\sigma=\dlk_n(\dlE)/\sqrt{2\dlE}$. The second part in (\ref{7.8}) then 
becomes
\begin{equation}
\label{7.9}
\int_{\dlE'}^{\dlmu}\rmd \dlE\, \sqrt{\dlmu-\dlE}\, 
\int_{-1}^{1}\rmd\sigma\,  
\rme^{\rmi \sqrt{2\dlE}\,\sigma\dly}\; \rho(\dlE,\sigma)
\end{equation}
with $\rho(\dlE,\sigma)\;\rmd\sigma$ the number of values of $\dlk_n(\dlE)$ 
in the interval $[\sqrt{2\dlE}\sigma,\sqrt{2\dlE}(\sigma+\rmd\sigma)]$. For 
large $\abs{\dly}$  the dominant contribution in the integral over $\sigma$ comes 
from the endpoints. Since the density $\rho(\dlE,\sigma)$ at the endpoints 
$\sigma=\pm 1$ is $2^{3/2}\dlE(1\mp\sigma)^{1/2}/\pi$ (see the appendix), 
the expression (\ref{7.9}) is for large $\abs{\dly}$
\begin{equation} 
\label{7.10}
\frac{2^{3/4}}{\pi^{1/2}\,{\abs{\dly}}^{3/2}}
\int_{\dlE'}^{\dlmu}\rmd \dlE\, \sqrt{\dlmu-\dlE}\; \dlE^{1/4}\, 
\cos\big(\sqrt{2\dlE}\, \abs{\dly}-3\pi/4\big).
\end{equation}
The integral over $\dlE$ can likewise be evaluated for large $\abs{\dly}$, as 
once again only the endpoints of the integral contribute. The upper boundary 
gives
\begin{equation}
\label{7.11}
  - \sqrt{2}\,\frac{\dlmu}{\dly^3}\,  \sin\big(\sqrt{2\dlmu}\, \dly\big)
\end{equation}
while the contribution from the lower boundary in (\ref{7.10}) will not be 
needed.

As a final step we have to consider the first part in (\ref{7.8}). Here the 
summation can be replaced by an integration for $\dlE\gg 1$ only. Of course, 
the contribution of the upper boundary to the asymptotic expression has to 
cancel that of the lower boundary in (\ref{7.10}), since the final result 
should not depend on the choice of $\dlE'$. The lower boundary of the integral 
over $\dlE$ in the first part of (\ref{7.8}) does not contribute either. In 
fact, only the term with $n=0$ survives for $\dlE$ near $1/2$ and the factor 
$\dlk_0(\dlE)$ goes to infinity for $\dlE \rightarrow 1/2$, so that for large 
$\abs{\dly}$ the contribution from the lower boundary damps out by 
interference.

Collecting the results, we have found that for $B\rightarrow 0$ and large 
$\abs{\dly}$ the asymptotic behaviour of the $\mu$-dependent Green function 
near the wall is given by (\ref{7.5}) with (\ref{7.11}) inserted:
\begin{equation}
G_\mu(xyz,xy'z) \approx
- \frac{4\,B^{3/2}\,\dlmu\,\dlx^2}{\pi^2\, \dly^3}\, 
 \sin\big(\sqrt{2\dlmu}\, \dly\big).
\label{7.12}
\end{equation}
After restoring the field-independent variables this result coincides with 
(\ref{7.7}). 

\subsection{Case $z\neq z'$}
In this case the dependence on $\dlk_n(\dlE)$ drops out from (\ref{7.4}):
\begin{equation}
\label{7.13}
\fl   G_\mu(xyz,xyz') \approx
  \frac{2B^{3/2}\, \dlx^2}{\pi^2 }\; 
  {\sum_n}'
  \int_{n+1/2}^{\dlmu} \rmd \dlE\;
  \frac{\sin\big[\sqrt{2[\dlmu-\dlE_n(\dlk)]}\, \dlz\big]}{\dlz}.
\end{equation}
The integral is elementary. As a result we get for the Green function at small 
$x$
\begin{eqnarray}           
\label{7.14}
  \fl G_\mu(xyz,xyz') \approx
-\frac{2 B^{3/2}\, \dlx^2}{\pi^2\dlz^3}\; 
{\sum_n}'\; \left\{\sqrt{2[\dlmu-(n+1/2)]}\, \dlz\, 
\cos\big[\sqrt{2[\dlmu-(n+1/2)]}\, \dlz\big]
\right.\nonumber\\
\left.-\sin\big[\sqrt{2[\dlmu-(n+1/2)]}\, \dlz\big]\right\}.
\end{eqnarray}
By taking the limit $\dlz\rightarrow 0$ we obtain a simple form for the 
particle density near the wall:
\begin{equation} 
\label{7.15}
\rho_{\mu}= G_\mu(\vect{r},\vect{r}) \approx
\frac{2^{5/2} B^{3/2}\, \dlx^2}{3\pi^2}\; 
  {\sum_n}'\; \left[ \dlmu-(n+1/2)\right]^{3/2}
\end{equation}    
which should be compared to the bulk density (\ref{3.7}).

For large separations the first term between the curly brackets in 
(\ref{7.14}) is dominant, so that the $\mu$-dependent Green function for large 
separations and small distance from the wall becomes
\begin{equation}
\label{7.16}
\fl G_\mu(xyz,xyz') \approx -\frac{2 B^{3/2}\, \dlx^2}{\pi^2 \dlz^2}
  {\sum_{n}}' \sqrt{2[\dlmu-(n+1/2)]}
  \cos\big[\sqrt{2[\dlmu-(n+1/2)]}\, \dlz\big].
\end{equation}
The same expression can also be found from (\ref{6.20}), by putting $\dlx\ll 1$ 
and expanding the hyperbolic function for small values of its argument.

For small $B$ it is convenient to return to the original variables $x$, 
$z-z'$, $\mu$, and to introduce the new integration variable $u=\dlE B$ in 
(\ref{7.13}). Replacing in addition the summation by an integration we get
\begin{equation}
\label{7.17}
  G_\mu(xyz,xyz') \approx
\frac{2 x^2}{\pi^2}\int_0^{\mu} \rmd t \int_t^{\mu} \rmd u \, 
\frac{\sin\big[\sqrt{2(\mu-u)}\, (z-z')\big]}{z-z'}.
\end{equation}
Performing the integrations we obtain the expression
\begin{eqnarray}
\label{7.18}
\fl  G_\mu(xyz,xyz') \approx
-\frac{2\mu x^2}{\pi^2 (z-z')^3}\left\{2\sin\big[\sqrt{2\mu}\,(z-z')\big]
+\frac{6}{\sqrt{2\mu}\,(z-z')}\cos\big[\sqrt{2\mu}\,(z-z')\big]\right.\nonumber\\
\left. -\frac{3}{\mu(z-z')^2}\sin\big[\sqrt{2\mu}\, (z-z')\big]\right\}
\end{eqnarray}
for the $\mu$-dependent Green function at positions near the wall in the 
field-free case. It can easily be checked that the same expression is found by 
expanding the general form (\ref{4.2}) for small $x$. 

At large separations $\abs{z-z'}$ the dominant term in (\ref{7.18}) is the first 
one. It agrees with (\ref{7.7}), when $y$ and $z$ are interchanged.

\section{Discussion and conclusion}

Our results show that the presence of a magnetic field and of a wall leads 
to remarkable changes in the pair correlation function of a completely 
degenerate non-interacting electron gas. In the bulk the correlation 
function follows from (\ref{2.7}), with the $\mu$-dependent Green functions 
(\ref{3.6}) and (\ref{3.14}) for the cases with and without field. Whereas 
the correlation function is isotropic and decays algebraically ($\propto 
r^{-4}$) in the field-free case, it becomes anisotropic with a Gaussian 
dependence in the transverse direction and an algebraic one ($\propto 
r^{-2}$) in the parallel direction. 

In the neighborhood of a plane hard wall the pair correlation function becomes 
anisotropic even for the field-free case. Its form, which follows directly 
from a reflection principle, is given by (\ref{2.7}) with (\ref{4.2}). For two 
positions far apart, but at equal distance from the wall, the ensuing 
correlation function decays algebraically, as in the bulk. However, the decay 
is proportional to $r^{-6}$, as follows from (\ref{7.7}). 

If in addition to the wall a magnetic field is present, the pair correlation 
function in the vicinity of the wall becomes anisotropic for two reasons, as 
both the field and the wall break the symmetry. The general expression for the 
correlation function, for arbitrary field strengths and arbitrary distances 
from the wall, follows by substitution of (\ref{5.7}) in (\ref{2.7}). 
Relatively far from the wall the corrections to the bulk correlation function 
are still small. The first-order correction terms were given in (\ref{4.14}) 
and (\ref{4.19}), and checked numerically in figures \ref{fig3} and \ref{fig5}. 
Near the wall the correlation function -- and in particular its tail for large 
separations -- is modified considerably as compared to its form in the bulk. 
In fact, the qualitative difference between the behaviour in the directions 
parallel and transverse to the field, which is a prominent feature of the bulk 
correlation function, is lost in the vicinity of the wall. In both directions 
the tails become algebraic, albeit with a different exponent. This is seen by 
inspecting (\ref{6.5}) (or (\ref{7.6})) for the transverse direction and 
(\ref{6.20}) (or (\ref{7.16})) for the parallel direction. In the former case 
the decay is proportional to $r^{-3}$, whereas in the latter it is $\propto 
r^{-4}$. Qualitatively, the change in the decay of the transverse correlation 
function from Gaussian in the bulk to algebraic near the wall can be 
understood in a semiclassical picture. In the bulk the cyclotron motion of the 
particles leads to a strong localization of the correlations, which is 
associated with a Gaussian decay. On approaching the wall the so-called 
`skipping orbits' along the wall become important. They lead to a 
delocalization effect in the particle motion, which manifests itself as an 
increase in the range of the correlations. In this way, the cross-over to an 
algebraic decay of the correlation function finds an explanation.

In conclusion, we have shown that for a non-interacting electron gas the 
influence of a wall on the correlations is quite considerable, especially in 
the presence of a magnetic field. It would be interesting to determine the 
edge effects in the correlations for an interacting electron gas in a magnetic 
field. 

\appendix
\section{Zeros of the parabolic cylinder functions for large argument and 
index}
The zeros of $D_{\lambda}(z)$ satisfy the inequality $\abs{z}<2\sqrt{\lambda+1/2}$, with 
the possible exception of a single negative zero \cite{ABR:1972}. For $\lambda\gg 
1$ the zeros in the neighborhood of $\abs{z}=2\sqrt{\lambda+1/2}$ can be determined by 
approximating the parabolic cylinder functions by Airy functions. 

For positive $z$ and $\sigma=z/[2\sqrt{\lambda+1/2}]$ slightly smaller than 1, 
one has \cite{ABR:1972}
\begin{equation}
\label{A1}
\fl
D_{\lambda}(z)\approx 2^{\lambda/2+1/3}\, \Gamma(\case{1}{2}\lambda+\case{1}{2})
(\lambda+\case{1}{2})^{1/6}\left(\frac{-\tau}{1-\sigma^2}\right)^{1/4}\;
{\rm Ai}\, \big([4(\lambda+\case{1}{2})]^{2/3}\tau\big)
\end{equation}
with $\tau$ given by
\begin{equation}
\label{A2}
\tau=-\left\{\case{3}{2}
\big[\case{1}{4}\arccos(\sigma)-\case{1}{4}\sigma\sqrt{1-\sigma^2}\big]\right\}^{2/3}
\approx - 2^{-1/3}(1-\sigma).
\end{equation}
For $x\gg 1$ the Airy function may be approximated as \cite{ABR:1972}
\begin{equation}
\label{A3}
{\rm Ai}(-x)\approx \frac{1}{\sqrt{\pi}\, x^{1/4}}\, \sin\left(\case{2}{3}x^{3/2}
+\case{\pi}{4}\right).
\end{equation}
Using this approximation in (\ref{A1}), one finds the zeros of the parabolic 
cylinder function from the zeros of the sine function. In terms of the 
variable $\sigma$, the zeros near $\sigma=1$ are found as
\begin{equation}
\label{A4}
\sigma_m=1-\frac{1}{2}\left[\frac{3\pi (m-1/4)}{2(\lambda+1/2)}\right]^{2/3}
\end{equation}
with positive integer $m$. Since $\lambda$ is large, the zeros are closely 
spaced. In view of the main text we introduce $\varepsilon=\lambda+1/2$ 
instead of $\lambda$. Writing the number of zeros between $\sigma$ and 
$\sigma+\rmd\sigma$ as $\rho(\varepsilon,
\sigma)\, \rmd\sigma$ we get
\begin{equation}
\label{A5}
\rho(\varepsilon,\sigma)=\frac{2^{3/2}}{\pi}\, \varepsilon\,(1-\sigma)^{1/2}
\end{equation}
as the density of zeros near $\sigma=1$. 

Similarly, for negative $z$ and $\sigma$ slightly bigger than $-1$ the 
parabolic cylinder function can be written as \cite{ABR:1972}
\begin{eqnarray}
\label{A6}
\fl
D_{\lambda}(z)\approx 2^{\lambda/2+1/3}\, \Gamma(\case{1}{2}\lambda+\case{1}{2})
(\lambda+\case{1}{2})^{1/6}\left(\frac{-\tau}{1-\sigma^2}\right)^{1/4}\nonumber\\
\times\left\{ \cos(\pi\lambda)\, {\rm Ai}\, \big([4(\lambda+\case{1}{2})]^{2/3}\tau\big)
-\sin(\pi\lambda)\, {\rm Bi}\, \big([4(\lambda+\case{1}{2})]^{2/3}\tau\big)\right\}
\end{eqnarray}
with $\tau\approx - 2^{-1/3}(1+\sigma)$. With the use of (\ref{A3}) and the 
analogous relation
\begin{equation}
\label{A7}
{\rm Bi}(-x)\approx \frac{1}{\sqrt{\pi}x^{1/4}}\, \cos\left(\case{2}{3}x^{3/2}
+\case{\pi}{4}\right)
\end{equation}
(valid for large $x$), one obtains the zeros of $D_{\lambda}(z)$ near $\sigma=-1$ as
\begin{equation}
\label{A8}
\sigma_m=-1+\frac{1}{2}\left[\frac{3\pi (m+\lambda-1/4)}{2(\lambda+1/2)}\right]^{2/3}
\end{equation}
for integer $m$ with $m\geq -\lambda+1/4$. The density of the zeros is found 
to be 
\begin{equation}
\label{A9}
\rho(\varepsilon,\sigma)=\frac{2^{3/2}}{\pi}\, \varepsilon \,(1+\sigma)^{1/2}
\end{equation}
for $\sigma$ near $-1$.


\section*{References}
\begin{thebibliography}{20}
\bibitem{BOH:1911} Bohr N 1911 {\em PhD Thesis}, in: 1972 {\em Collected Works 
I} (Amsterdam: North-Holland); Van Leeuwen H.J.\ 1919 {\em PhD Thesis}, 1921 
\journaltitle{J.\ de Physique} \volume{2} 361; Landau L 1930 \journaltitle{Z.\ 
Phys.} \volume{64} 629

\bibitem{OHT:197397} Ohtaka K and Moriya T 1973 \journaltitle{J.\ Phys.\ 
Soc.\ Japan} \volume{34} 1203; Jancovici B 1980 \journaltitle{Physica} 
\volume{101A} 324; Macris N, Martin Ph A and Pul\'e J V 1988 
\journaltitle{Comm.\ Math.\ Phys.} \volume{117} 215; Macris N, Martin Ph A  
and Pul\'{e} J V 1997 \journaltitle{Ann.\ Inst.\ H.\ Poincar\'{e}} 
\volume{66} 147

\bibitem{JOH:1994} John P and Suttorp L G 1994 \journaltitle{Physica A} 
\volume{210} 237

\bibitem{JOH:1995} John P and Suttorp L G 1995 \journaltitle{J.\ Phys.\ A: 
Math.\ Gen.} \volume{28} 6087

\bibitem{KES:1998} Kettenis M M and Suttorp L G 1998 \journaltitle{J.\ Phys.\ 
A: Math.\ Gen.} \volume{31} 6547

\bibitem{KES:1999} Kettenis M M and Suttorp L G 1999 \journaltitle{J.\ Phys.\ 
A: Math.\ Gen.} \volume{32} 8209

\bibitem{ISI:1971} 
Isihara A, Tsai J and Wadati M 1971 \journaltitle{Phys.\ Rev.\ A} \volume{3} 990

\bibitem{ALA:198897} Alastuey A and Martin Ph A 1988 \journaltitle{Europhys.\ 
Lett.} \volume{6} 385; 1989 \journaltitle{Phys.\ Rev.\ A} \volume{40} 6485; 
Cornu F and Martin Ph A 1991 \journaltitle{Phys.\ Rev.\ A} \volume{44} 4893; 
Cornu F 1996 \journaltitle{Phys.\ Rev.\ E} \volume{53} 4562, 4595; Alastuey A 
and Cornu F 1997 \journaltitle{J.\ Stat.\ Phys.} \volume{89} 20; Cornu F 1997 
\journaltitle{Phys.\ Rev.\ Lett.} \volume{78} 1464  

\bibitem{COR:199798} Cornu F 1997 \journaltitle{Europhys.\ Lett.} \volume{37} 
591; 1998 \journaltitle{Phys.\ Rev.\ E} \volume{58} 5268, 5322

\bibitem{ROEP:1996} Roepstorff R 1994 \booktitle{Path Integral Approach to 
Quantum Physics} (Berlin: Springer)

\bibitem{AQU:1999} Aqua J N and Cornu F 1999 \journaltitle{J.\ Stat.\ Phys.} 
\volume{97} 173

\bibitem{JAN:1985} Jancovici B 1985 \journaltitle{J.\ Stat.\ Phys.} 
\volume{39} 427; Jancovici B, Lebowitz J L and Martin Ph A 1985 
\journaltitle{J.\ Stat.\ Phys.} \volume{41} 941

\bibitem{SOW:1951} Sondheimer E H and Wilson A H 1951
  \journaltitle{Proc.\ R.\ Soc.\ A} \volume{210} 173

\bibitem{GIN:1971} Ginibre J 1971 {\em in:} DeWitt C and Stora R eds., {\em 
Statistical Mechanics and Quantum Field Theory} (New York: Gordon and Breach) 
p.\ 329

\bibitem{MOS:1966} Magnus W, Oberhettinger F, and Soni R P 1966
  {\em Formulas and Theorems for the Special Functions of Mathematical
    Physics} (Berlin: Springer)

\bibitem{ERD:1954}  Erd\'elyi A (ed.) 1954 {\em Tables of Integral 
Transforms vol.~I} (New York: McGraw-Hill) p.\ 245 

\bibitem{ERD:1954a}  Erd\'elyi A (ed.) 1954 {\em Tables of Integral Transforms 
vol.~II} (New York: McGraw-Hill) p.\ 335

\bibitem{AUK:1985} Auerbach A and Kivelson S 1985 \journaltitle{Nucl.\ Phys. 
B} \volume{257} 799

\bibitem{BAB:1970} Balian R and Bloch C 1970 \journaltitle{Ann.\ Phys., NY} 
\volume{60} 401

\bibitem{ERD:1956} Erd\'elyi A  1956 {\em Asymptotic expansions} (New York: 
Dover) p.\ 51

\bibitem{WON:1989} Wong R 1989 {\em Asymptotic approximations of integrals} 
(Boston: Academic Press) p.\ 74

\bibitem{ABR:1972} Abramowitz M and Stegun I A 1972 {\em Handbook of 
Mathematical Functions} (New York: Dover) Ch.\ 19

\end{thebibliography}
\end{document}